\documentclass[journal]{IEEEtran}
\usepackage{amsthm}
\usepackage{inputenc}
\usepackage{ifpdf}
\usepackage{xcolor}
\usepackage{cite}
\usepackage{amsmath}
\usepackage{siunitx}
\usepackage{array}
\usepackage{graphicx}
\usepackage{url}
\usepackage{tabularx}
\usepackage[english]{babel}
\usepackage{subcaption}
\usepackage{multirow}
\usepackage{hhline}
\usepackage{amsmath,amssymb}
\usepackage{multirow}
\usepackage{caption}
\usepackage{amssymb}
\usepackage{graphicx}
\usepackage{epstopdf}
\usepackage{cite}
\usepackage{siunitx}
\usepackage{color}
\usepackage{amssymb}

\usepackage{amsthm}
\usepackage{tikz}
\usepackage{mathtools}
\usepackage{siunitx}

\usepackage{listings}
\usepackage{eurosym}
\usepackage{footnote}
\usepackage{booktabs,caption}
\usepackage{pifont}

\usepackage{amsmath}

\newcommand{\bDiamond}{\mathbin{\Diamond}}

\DeclareSIUnit[inter-unit-product={}] \MVA {\mega\volt\ampere} %apparent power
\DeclareSIUnit[inter-unit-product={}] \MWh {\MW\hour} %apparent power
\DeclareSIUnit \pu {p.u.}
\DeclareSIUnit[inter-unit-product={}] \KV {\kilo\volt} %KV

\newtheorem{Theorem}{Theorem}

\newtheorem{Corollary}{Corollary}

\DeclareSIUnit[inter-unit-product={}] \MVA {\mega\volt\ampere} %apparent power
\DeclareSIUnit[inter-unit-product={}] \MWh {\MW\hour} %apparent power
\DeclareSIUnit \pu {p.u.}
\DeclareSIUnit[inter-unit-product={}] \KV {\kilo\volt} 

\begin{document}
	\setlength{\abovedisplayskip}{2pt}
\setlength{\belowdisplayskip}{2pt}
\setlength{\textfloatsep}{2pt}
\title{A Scenario-based Branch-and-Bound Approach for MES Scheduling in Urban Buildings}

\author{Mainak~Dan,~\IEEEmembership{Student Member}, Seshadhri~Srinivasan,~\IEEEmembership{Sr. Member}, Suresh Sundaram,~\IEEEmembership{Sr. Member}, Arvind Easwaran,~\IEEEmembership{Sr. Member} and Luigi Glielmo,~\IEEEmembership{Sr. Member} 
        % <-this % stops a space
 \thanks{Mainak Dan is with the Interdisciplinary Graduate Program, Nanyang Technological University, Singapore. e-mail: mainak001@e.ntu.edu.sg.}
 \thanks{Seshadhri Srinivasan is with Berkeley Education Alliance for Research in Singapore (BEARS). e-mail : seshucontrol@gmail.com.}
 \thanks{Suresh Sundaram is with the Department of Aerospace Engineering, Indian Institute of Science, Bangalore, India. e-mail : vssuresh@iisc.ac.in.}
 \thanks{Arvind Easwaran is with the School of Computer Science and Engineering, Nanyang Technological University, Singapore. e-mail : arvinde@ntu.edu.sg.}
 \thanks{Luigi Glielmo is with the Department of  Engineering, University of Sannio, Benevento, Italy. e-mail:glielmo@unisannio.it}
}

\maketitle
\begin{abstract}
This paper presents a novel solution technique for scheduling multi-energy system (MES) in a commercial urban building to perform price-based demand response and reduce energy costs. The MES scheduling problem is formulated as a mixed integer nonlinear program (MINLP), a non-convex NP-hard problem with uncertainties due to renewable generation and demand. A model predictive control approach is used to handle the uncertainties and price variations. This in-turn requires solving a time-coupled multi-time step MINLP during each time-epoch which is computationally intensive. This investigation proposes an approach called the Scenario-Based Branch-and-Bound (SB3), a light-weight solver to reduce the computational complexity. It combines the simplicity of convex programs with the ability of meta-heuristic techniques to handle complex nonlinear problems. The performance of the SB3 solver is validated in the Cleantech building, Singapore and the results demonstrate that the proposed algorithm reduces energy cost by about 17.26\% and 22.46\% as against solving a multi-time step heuristic optimization model.
\end{abstract}

\IEEEpeerreviewmaketitle

\section*{Nomenclature} \vspace{-15pt}
\noindent \subsection*{Dispatchable Generators} \vspace{-4pt}
\begin{IEEEdescription}[\IEEEusemathlabelsep\IEEEsetlabelwidth{$P_{\rm{loss}}$~ }]
\item[$P_{{\rm{G}},i}$] Power generated by the generator $i$ [\si{\kilo\watt}];
\item[$\delta_{{\rm{G}},i}$] Binary ON-OFF status of the generator $i$;
\item [$R_{{\rm{G}},i}$] Ramp-rate of the generator [\si{\kilo\watt\per\hour}];
\end{IEEEdescription}\vspace{-10pt}
\subsection*{Cogeneration Units}\vspace{-4pt}
\begin{IEEEdescription}[\IEEEusemathlabelsep\IEEEsetlabelwidth{$P_{\rm{loss}}$~ }]
\item[$P_{\rm{GT}}$] Power generated by the gas-turbine [\si{\kilo\watt}];
\item[$\delta_{\rm{GT}}$] Binary ON-OFF status of the gas-turbine;
\item [$R_{{\rm{GT}}}$] Ramp-rate of the gas-turbine [\si{\kilo\watt\per\hour}];
\item[$Q_{\rm{AC}}$] Thermal output of the absorption chiller [\si{\kilo\watt}];
\end{IEEEdescription}\vspace{-10pt}
\subsection*{Utility Grid and Renewable Energy}\vspace{-4pt}
\begin{IEEEdescription}[\IEEEusemathlabelsep\IEEEsetlabelwidth{$P_{\rm{loss}}$~ }]
\item[$P_{\rm{GR}}$] Power bought from the grid [\si{\kilo\watt}];
\item[$P_{\rm{R}}$]  Power generated by the photo-voltaic panel [\si{\kilo\watt}];
\end{IEEEdescription}\vspace{-10pt}
\subsection*{Storage Units}\vspace{-4pt}
\begin{IEEEdescription}[\IEEEusemathlabelsep\IEEEsetlabelwidth{$P_{\rm{loss}}$~ }]
\item[$P_{\rm{ESS}}$]  Power input to the electrical storage [\si{\kilo\watt}];
\item [${SoC}_{\rm{ESS}}$] State of charge of the electrical storage [\si{\kilo\watt\hour}];
\item [$\delta_{\rm{ESS}}$] charging/discharging status of the electrical storage;
\item[$P_{\rm{TES}}$]  Cooling energy input to the thermal storage [\si{\kilo\watt}];
\item [${SoC}_{\rm{TES}}$] State of charge of the thermal storage [\si{\kilo\watt\hour}];
\item [$\delta_{\rm{TES}}$] charging/discharging status of the thermal storage;
\item [$T_{\rm{TES}}$]  Temperature inside the TES [\si{\celsius}];
\end{IEEEdescription}\vspace{-10pt}
\subsection*{Chiller Bank}\vspace{-4pt}
\begin{IEEEdescription}[\IEEEusemathlabelsep\IEEEsetlabelwidth{$P_{\rm{loss}}$~ }]
\item[$\dot{m}_{{\rm{C}},j}$] Cool water mass-flow rate of chiller $j$ [\si{\kilogram\per\hour}];%(kg/s)
\item[$\gamma_{{\rm{C}},j}$] Binary ON-OFF status of chiller $j$;	
\item[$P_{\rm{C}}$]  Power consumed in the chiller bank [\si{\kilo\watt}];
\item [$Q_{{\rm{C}},j}$] Thermal energy supplied by chiller $j$ [\si{\kilo\watt}];
\end{IEEEdescription}\vspace{-10pt}
\subsection*{Energy Demands}\vspace{-4pt}
\begin{IEEEdescription}[\IEEEusemathlabelsep\IEEEsetlabelwidth{$P_{\rm{loss}}$~ }]
\item[$P_{\rm{L}}$] Electrical load demand [\si{\kilo\watt}];
\item[$Q_{\rm{L}}$] Thermal load demand [\si{\kilo\watt}];
\end{IEEEdescription}\vspace{-10pt}
\subsection*{Parameters}\vspace{-4pt}
\begin{IEEEdescription}[\IEEEusemathlabelsep\IEEEsetlabelwidth{$P_{\rm{loss}}$~ }]
\item[$H_L$] Coefficient of heat loss [\si{\watt\per\meter\squared\per\celsius}];
\item [$T_{\rm{in/out}}$]  Inlet/Outlet temperature of the TES [\si{\celsius}];
\item[$V$]  Volume of the TES [\si{\meter\cubed}];
\item[$A$]  Cross-sectional area of the TES [\si{\meter\squared}];
\item[$C_{\rm{GR}}$]  Market energy cost [\si{\$\per\kilo\watt\hour}];
\item [$\Delta t$] Sampling time [\si{\hour}];
\item [$\eta_{\rm{ESS}}$] Electrical storage efficiency;
\item [$\eta_{\rm{TES}}$] Thermal storage efficiency;
%\item [$n_c$] Number of chillers;
%\item [$n_g$] Number of generators;
%\item [$\theta^{\rm{CHP}}_{{te}}$] Conversion factor for CHP thermal to electrical;
\item [$C_{\rho}$] Specific heat capacity of the water [\si{\kilo\joule\per\kilogram\per\celsius}];
\item [$\rho$] Density of the water [\si{\kilogram\per\meter\cubed}];
\item [$k$] Sampling index;
\item [$N_p$] Time horizon;
\end{IEEEdescription}

\IEEEpeerreviewmaketitle

\section{Introduction}

\IEEEPARstart{T}{he} multi-energy systems (MES) is an emerging concept wherein different energy vectors (e.g., thermal, electrical, gas and co-generation units) optimally interact at various levels in smart grid environments~\cite{Clegg2016}. It is increasingly being recognized that a well-coordinated scheduling of different energy systems can reduce the energy cost, increase overall efficiency, mitigate peak-demand and guarantee reliable power delivery~\cite{sheikhi2016}. Key challenge here lies in optimally integrating different energy vectors and networks~(e.g., thermal and electric)~\cite{mavromatidis2017}. However, the presence of tight coupling between multiple energy vectors, components' complex behaviours (e.g., nonlinear, switching, and time-coupled), and many degrees of freedom make the optimal scheduling problem intricate. The intermittent renewable generation and pulsating demand add a new dimension to the problem. Therefore, novel control approaches are required for scheduling MES.

Optimization models, scheduling techniques, and control approaches are widely investigated problems in energy management system (EMS) related to MES. The energy hub (EH) concept proposed in~\cite{kienzle2011} is a widely adopted model for studying MES energy exchanges. The optimization approaches for solving such models in the literature are categorised as: {\em{(i)}} mathematical programming~\cite{ameri2016,mancarella2013},  {\em{(ii)}} meta-heuristic~\cite{bao2015a}, {\em{(iii)}} stochastic programming~\cite{dolatabadi2018} and {\em{(iv)}} hybrid approach~\cite{morvaj2017}. 
Mathematical models for MES scheduling include the mixed integer linear programming (MILP)~\cite{sudtharalingam2012micro,de2013optimal,Verrilli2017,shao2017,pal2018electric,jindal2018heuristic}, which inherently fits the MES operating modes and is widely used tool due to its ability to include switching actions. Oversimplification of nonlinear behaviours by incorporating linear relaxation and the lack of scalability with the number of integer constraints are the two shortcomings of the MILP models. The nonlinear programs~\cite{Catalao2017,moeini2014} and mixed integer nonlinear models (MINLP)~\cite{zheng2018,gabrielli2017} comprehensively capture the complex behaviours of the MES. However, due to the non-linearity and non-convexity ~\cite{mencarelli2017multiplicative,morvaj2017}, the existing commercial solvers for MINLP require more computation resources, solution time, and are sensitive to initial conditions~\cite{burer2012}. 

The meta-heuristic techniques are often used to solve non-convex MINLPs with single as well as multiple objectives. There are several existing literature which have proposed modifications of meta-heuristic techniques to improve the convergence rate while solving economic load dispatch problem of EMS~\cite{sun2013solving,bao2015a,askarzadeh2017memory,shefaei2017wild,reynolds2019operational}. However, their performances in non-convex MINLPs are dependent upon the quality of the initial solutions (seeds). As a result, the meta-heuristic approaches usually have large computation times, thereby making them unsuitable for real-time operations~\cite{Dan2018ssci}. More recently, hybrid approaches that combine mathematical programming and meta-heuristic techniques have been proposed~\cite{morvaj2017}. Though combined methods provide a sub-optimal solution, they solve complex problems in reasonable time and have the potential to address the non-linearity encountered in dispatch problems of smart-grid environments with distributed multi-energy sources. Still the aforementioned works have not considered intermittent behaviours in their analysis. 

Two widely used methods for handling intermittent behaviours are stochastic programming (SP)~\cite{dolatabadi2018,huo2019chance,zhao2019two} and model predictive control (MPC)~\cite{Verrilli2017,maffei2018}. Typically SP solves a set of scenarios envisaged due to fluctuating behaviours from normal conditions~\cite{dolatabadi2018,cao2019networked}. This requires solving multiple instances of MINLP model that leads to scalability issues. The MPC on the other hand solves multi-time step problem considering the underlying disturbances and encapsulating complex behaviours. This boils down to a complex optimization models. In addition, implementing MPC requires specialized hardware with commercial solvers which are cost-intensive. A review of the literature reveals that MINLP models capture the complex MES behaviours\cite{zheng2018,gabrielli2017} and MPC approach is more suited for handling uncertainties. Although commercial solvers are available for solving MINLP problems, constraints on computational resources and light-weight solvers that can be ported into EMS control hardware are very much in demand and more often when MPC approach is used, the optimal solution is guaranteed from only relatively small-scale convex MINLPs. Emerging multi-time step MINLPs in EMS with large number of continuous and binary decision variables as well as non-convex constraints represent a difficult challenge and finding a feasible solution is computationally challenging \cite{liberti2011recipe}. Consequently, efficient solvers with fast convergence rate for MINLP model based MPC are required.

This investigation designs a MPC for scheduling MES devices to perform price-based demand response (PBDR) and reduce operating cost. The scheduling problem is formulated as a non-convex MINLP-MPC model. In order to solve the problem, a novel light-weight solution method called the scenario-based branch-and-bound (SB3) is proposed. The SB3 is  hybrid MINLP solver that integrates convergence efficiency of convex mathematical programming as well as the exhaustive searching capability of meta-heuristic solver. The main contributions of this study are: {\em{(i)}} proposition of a light-weight solver called the scenario-based branch-and-bound (SB3) for scheduling MES devices, {\em{(ii)}} investigations into the existence conditions of the lower bound solution of the SB3 algorithm, and {\em{(iii)}} demonstration of the SB3 approach in a pilot building, which is a community microgrid and is responsible for supplying energy to the neighbouring buildings and office spaces.
In order to understand the efficiency, performance of the SB3 algorithm is  compared with performance of other MINLP solvers.

\begin{figure}[htb]
\centering
\includegraphics[height=5.0cm,width=9cm]{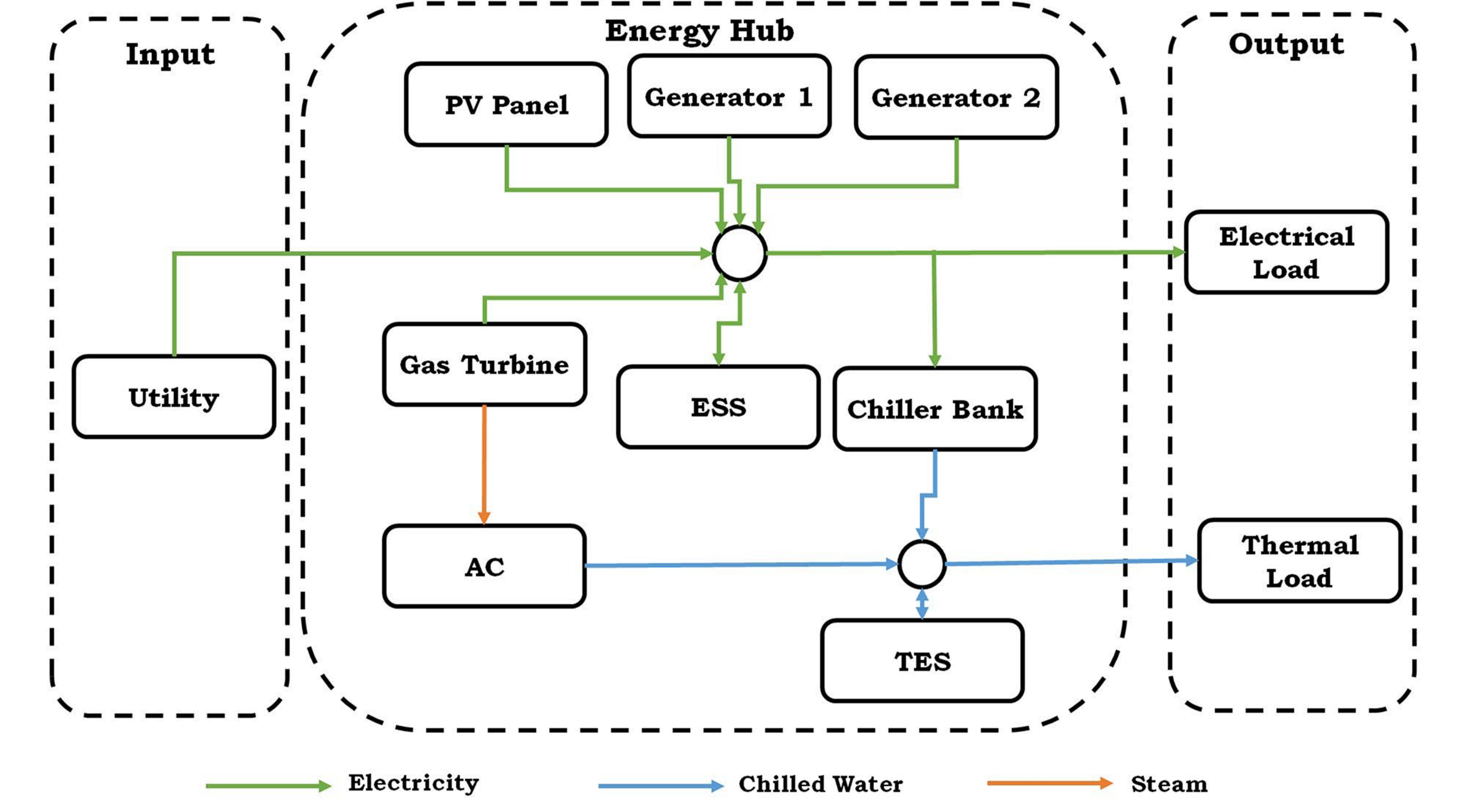}
\caption{Multi-Energy Systems Architecture of Cleantech Building, Singapore} 
\label{fig:MES}
\end{figure}
The paper is organized as follows. Section~\ref{sec:MESModel} presents the MES component models and the optimization problem. Section~\ref{sec:SB3} describes the proposed SB3 algorithm and its different components. with a mathematical analysis. The results are provided in Section~\ref{Sec:Results} and conclusions in Section \ref{Sec:Concl}.

\section{MES Model and Problem Formulation} \label{sec:MESModel}

The energy hub model of the MES studied in this investigation is shown in Fig.~\ref{fig:MES}. The supply from utility is the input to source energy demand. In addition, the electrical energy is supplied by two dispatchable generators (DG), one gas turbine (GT) embedded within co-generation unit (CGU), and a photo-voltaic panel. Similarly, thermal energy demand is served by the five chillers and one absorption chiller (AC) that resides inside the CGU. Steam generated from GT water-jacket is used to source the AC through heat recovery and system generator. The EH also includes electrical storage system (ESS) and thermal energy storage (TES) for handling intermittent behaviours.

\subsection{Generator Operating Constraints (DG)}

Suppose that the state of the generator is denoted by $\delta_{{\rm{G}},i}^k$ and the minimum up/down time by  \small$T_{{\rm{G}},i}^{\rm{up}}$\normalsize~and \small$T_{{\rm{G}},i}^{\rm{down}}$\normalsize,~i.e. the minimum time the unit should stay in a particular state before toggling to the other state. Thus, the temporal constraints associated with the generator during each sampling time $k$ are modelled as \cite{Verrilli2017}

\small
\begin{subequations} \label{G1}
\begin{align}
\delta_{{\rm{G}},i}^k-\delta_{{\rm{G}},i}^{k-1} \le \delta_{{\rm{G}},i}(\tau_{{\rm{G}},i}^{\rm{up}}),~~~~~\label{subeq1}\\
\delta_{{\rm{G}},i}^{k-1}-\delta_{{\rm{G}},i}^k \le 1-\delta_{{\rm{G}},i}(\tau_{{\rm{G}},i}^{\rm{down}}),\label{subeq2}
\end{align}
\end{subequations}
\normalsize

\noindent where $i \in \{1, \dots, n_g\}$ is the index of the generator, $n_g$ is the number of generators, $\tau_{{\rm{G}},i}^{\rm{up}} = k+1, \dots, \operatorname{min}(k+T^{\rm{up}}_{{\rm{G}},i}-1, N_p) $, and $\tau_{{\rm{G}},i}^{\rm{down}} = k+1, \dots, \operatorname{min}(k+T^{\rm{down}}_{{\rm{G}},i}-1, N_p) $, respectively. Similarly, the generation limits and ramp-rate of the generators are modelled as,

\small
\begin{subequations} \label{G2}
\begin{align}
\delta_{{\rm{G}},i}^k~\underline{P}_{{\rm{G}},i} \leq P_{{\rm{G}},i}^k \leq \delta_{{\rm{G}},i}^k~\overline{P}_{{\rm{G}},i}~~~\label{subeq3}\\ 
|P_{{\rm{G}},i}^k-P_{{\rm{G}},i}^{k-1}| \le R_{{\rm{G}},i},~\forall i~\label{subeq4}
\end{align}
\end{subequations}
\normalsize

\noindent where $P_{{\rm{G}},i}^k$ and $R_{{\rm{G}},i}$ are the power provided by the generator and the ramp-rate limit of the generator $i$, and $\underline{P}_{{\rm{G}},i}$ and $\overline{P}_{{\rm{G}},i}$ denote lower and upper generation limits of the $i^{\rm{th}}$ generator. The generation cost is given by~\cite{Verrilli2017}

\small
\begin{equation}
C\left(P_{{\rm{G}},i}^k\right)~=~\sum_{i=1}^{n_g} \delta_{{\rm{G}},i}^k\left(A_{{\rm{G}},i}~{P_{{\rm{G}},i}^k}^2+ B_{{\rm{G}},i} {P_{{\rm{G}},i}^k} +C_{{\rm{G}},i}\right), \label{eq:Gencost}
\end{equation}
\normalsize

\noindent where $A_{{\rm{G}},i}$, $B_{{\rm{G}},i}$ and $C_{{\rm{G}},i}$ are coefficients. 
\subsection{Cogeneration Units (CGU)}
Similar to the generator, the constraints associated with GT's minimum up/down time are defined during each sampling time $k$ as:

\small
\begin{subequations} \label{CG1}
\begin{align}
\delta_{\rm{GT}}^k-\delta_{\rm{GT}}^{k-1} \le \delta_{\rm{GT}}(\tau_{\rm{GT}}^{\rm{up}})~~~~\label{subeq5}\\
\delta_{\rm{GT}}^{k-1}-\delta_{\rm{GT}}^k \le 1-\delta_{\rm{GT}}(\tau_{\rm{GT}}^{\rm{down}}),\label{subeq6}
\end{align}
\end{subequations}
\normalsize

\noindent where $\tau_{\rm{GT}}^{\rm{up}} = k+1, \dots, \operatorname{min}(k+T^{\rm{up}}_{\rm{GT}}-1, N_p) $, and $\tau_{\rm{GT}}^{\rm{down}} = k+1, \dots, \operatorname{min}(k+T_{\rm{GT}}^{\rm{down}}-1, N_p) $, respectively. The physical bounds and the ramp-rate constraints are modelled as

\small
\begin{subequations} \label{CG2}
\begin{align}
\delta_{\rm{GT}}^k\underline{P}_{\rm{GT}} \leq P_{\rm{GT}}^k \leq \delta_{\rm{GT}}^k \overline{P}_{\rm{GT}} \label{subeq8}\\
|P_{\rm{GT}}^k-P_{\rm{GT}}^{k-1}| \le R_{\rm{GT}},~~~\label{subeq7}
\end{align}
\end{subequations}
\normalsize

\noindent where $P_{\rm{GT}}^k$ is the power generated by the gas-turbine and $R_{\rm{GT}}$ models the ramp-rate constraint. In addition, $[\underline{P}_{\rm{GT}}, \overline{P}_{\rm{GT}}]$ denotes the generation limits of the gas-turbine whose cost is given by

\small
\begin{equation}
C\left(P_{\rm{GT}}^k\right) = \delta_{\rm{GT}}^k \left( A_{\rm{GT}} {P_{\rm{GT}}^k}^2+B_{\rm{GT}} P_{\rm{GT}}^k +C_{\rm{GT}} \right),
\end{equation}
\normalsize

\noindent where $A_{\rm{GT}},~B_{\rm{GT}}$ and $C_{\rm{GT}}$ are coefficients. To increase energy efficiency, the heat recovered by the water jacket of the GT is used to feed the absorption chiller along with boiler. Considering that the heat energy provided by the GT depends linearly on the power generated, we have

\small
\begin{equation} \label{eq:QACHC1}
{Q_{\rm{AC}}^{\rm{inp}}}^k= \beta_{\rm{GT}} P_{\rm{GT}}^k.
\end{equation}
\normalsize

\noindent Moreover, the ${Q_{\rm{AC}}^{\rm{inp}}}^k$ is related to the cooling energy supplied by the chiller using the equation

\small
\begin{equation} \label{eq:ACOP}
Q_{\rm{AC}}^k = {\rm{COP_{AC}}} {Q_{\rm{AC}}^{\rm{inp}}}^k =  {\rm{COP_{AC}}} \beta_{\rm{GT}} P_{\rm{GT}}^k,
\end{equation}
\normalsize

\noindent where ${\rm{COP_{AC}}}$ is the coefficient of performance of the AC for the given thermal load at time-instant $k$. Further, the cooling energy supplied by the AC is limited through the following constraints:

\small
\begin{equation}\label{CG4}
\delta_{\rm{GT}}^k\underline{Q}_{\rm{AC}} \le Q_{\rm{AC}}^k \le \delta_{\rm{GT}}^k \overline{Q}_{\rm{AC}},
\end{equation}
\normalsize

\noindent where $[\underline{Q}_{\rm{AC}}, \overline{Q}_{\rm{AC}}]$ are the limits of the thermal energy produced by the AC.

\subsection{Chiller Bank (CB)}
In addition to the AC, there is a CB consisting of five chillers whose mass-flow rates are controlled using a supervisory controller and the outlet temperature is maintained at a pre-defined set-point using closed-loop controls.  Following \cite{deng2014model}, the cooling energy supplied from the CB is given by

\small
\begin{equation}\label{eq:chiller1}
Q_{{\rm{C}},j}^k = \gamma_{{\rm{C}},j}^k \dot{m}_{{\rm{C}},j}^k C_\rho \Delta T_{{\rm{C}}},~~ \forall j \in \{1, \dots, n_c\},
\end{equation}  
\normalsize

\noindent where $\dot{m}_{{\rm{C}},j}^k$ is the mass-flow rate of the $j^{\rm{th}}$ chiller and $\gamma_{{\rm{C}},j}^k$ denotes the corresponding ON/OFF operation and $\Delta T_{{\rm{C}}}$ is the temperature difference between the inlet and outlet of the chiller. The electrical power consumed by CB is given by
\small
\begin{align}
P_{{\rm{C}}}^k =\sum_{j \in \{1, \dots, n_c\}} \frac{Q_{{\rm{C}},j}^k}{{\rm{COP}}_{{\rm{C}},j}}.
\end{align} 
\normalsize
\noindent In addition, the mass-flow is bounded by
\small
\begin{equation}\label{eq:chiller2}
\gamma_{{\rm{C}},j}^k \underline{\dot{m}}_{{\rm{C}},j} \le \dot{m}_{{\rm{C}},j}^k \le \gamma_{{\rm{C}},j}^k \overline{\dot{m}}_{{\rm{C}},j},
\end{equation}
\normalsize
\noindent where $\underline{\dot{m}}_{{\rm{C}},j}$ and $\overline{\dot{m}}_{{\rm{C}},j}$ denote the lower and upper bound on the chiller mass-flow rates.

\subsection{Energy Exchanged with the Utility Grid}

The power exchanged with the utility grid $P_{\rm{GR}}^k$ is constrained by lower ($\underline{P}_{\rm{GR}}$) and upper bound ($\overline{P}_{\rm{GR}}$) on the consumption given by
\small
\begin{equation}\label{eq:PU}
-\underline{P}_{\rm{GR}} \le P_{\rm{GR}}^k \le \overline{P}_{\rm{GR}}.
\end{equation}
\normalsize
\noindent The cost of power exchanged with the utility grid is time-varying and and modelled as,
\small
\begin{equation}
C_{\rm{GR}}^k = C_{\rm{D}}^k + C_{\rm{RT}}^k,
\end{equation}
\normalsize
\noindent where $C_{\rm{D}}^k$  and $C_{\rm{RT}}^k$ denote the day-ahead and real-time demand-based price of energy respectively at time instant $k$.

\subsection{Thermal Energy Storage (TES)}
\noindent Following \cite{celador2011}, the TES temperature dynamics is 
\small
\begin{align}\label{eq:TES}
 T_{\rm{TES}}^{k+1}=&~ \frac{\mu}{V \rho }\left(\sum_{j=1}^{n_c}\dot{m}_{{\rm{C}},j}^k \gamma_{{\rm{C}},j}^k \right) \left( T_{\rm{in}}^k-T_{\rm{TES}}^{k} \right)+T_{\rm{TES}}^{k} \left(1-\frac{H_L A}{V \rho C_{\rho}}\right)& \nonumber\\
&~~~~~~~~~~~~~~~~~~~~~~~~~~~~+T_{\rm{amb}}\left(\frac{H_L A}{V \rho C_{\rho}}\right),&\\
& ~~~~~~~~~~~~\underline{T}_{\rm{TES}} \leq T_{\rm{TES}}^{k} \leq \overline{T}_{\rm{TES}} \nonumber&, 
\end{align}
\normalsize

\noindent where, $T_{\rm{in}}^k,T_{\rm{TES}}^k$ and $T_{\rm{amb}}^k$ denote the inlet temperature of the TES, inlet temperature and the ambient temperature in \si{\celsius}. $0 < \mu < 1$ defines the fraction of chilled water shared between CB and TES. The energy stored in the TES is modelled as 

\small
\begin{subequations} \label{TES1}
\begin{align}
&SoC_{\rm{TES}}^{k+1}~=~ SoC_{\rm{TES}}^{k}+ \delta_{\rm{TES}}^{k}\eta^c_{\rm{TES}} P_{\rm{TES}}^{+^k}\Delta t&\label{subeq11}\\
&~~~~~~~~~~~~~~~~~~~~+(1-\delta_{\rm{TES}}^k)\frac{1}{\eta^d_{\rm{TES}}} P_{\rm{TES}}^{-^k} \Delta t+H_L A \Delta T,&\nonumber\\
&~~~~~~~~~~~~~~~~~~~\underline{SoC}_{\rm{TES}}\le {SoC}_{\rm{TES}}^k \le \overline{SoC}_{\rm{TES}},&\label{subeq12}
\end{align}
\end{subequations}
\normalsize

\noindent where ${SoC}_{\rm{TES}}^k$ and $P_{\rm{TES}}^{{+/-}^k}$ denote the thermal storage, the power exchanged (charging/ discharging) with the TES, respectively, and $\Delta T$ is the difference between the ambient and the TES's internal temperature. The binary variable $\delta_{\rm{TES}}^k$ models the charging/discharging operations and $0<\eta_{\rm{TES}}^c,\eta_{\rm{TES}}^d <1$. The storage power is also constrained by the physical limits of the TES which is given by

\small
\begin{equation}\label{eq:PTlast}
-(1-\delta^k_{\rm{TES}})~\underline{P}_{\rm{TES}} \le P_{\rm{TES}}^k \le \delta_{\rm{TES}}^k~ \overline{P}_{\rm{TES}},
\end{equation} 
\normalsize

\noindent where $[\underline{P}_{\rm{TES}},\overline{P}_{\rm{TES}}]$ denote the limits on $P_{\rm{TES}}$. 

\subsection{Electrical Storage Systems (ESS)}

The state-of-charge (SoC) of the ESS is modelled as\cite{maffei2018}:

\small
\begin{subequations}
\begin{align}
&SoC^{k+1}_{\rm{ESS}}=SoC^{k}_{\rm{ESS}}+ \delta_{\rm{ESS}}^k \eta^c_{\rm{ESS}} P_{\rm{ESS}}^{+^k}\Delta t 
&\nonumber\\
&~~~~~~~~~~~+(1-\delta_{\rm{ESS}}^k)\frac{1}{\eta^d_{\rm{ESS}}} P_{\rm{ESS}}^{-^k}\Delta t- \eta_{\rm{loss}}SoC^{k}_{\rm{ESS}}\label{subeq13}&\\
&~~~~~~~~~~~~~~~~~~~\underline{SoC}_{\rm{ESS}}\le {SoC}_{\rm{ESS}}^k \le \overline{SoC}_{\rm{ESS}},&
\end{align}
\end{subequations}
\normalsize

\noindent where ${SoC}^k_{\rm{ESS}}$ is state of charge, $P_{\rm{ESS}}^{+/-^k}$ is the power exchanged with storage (charging/ discharging) and $\delta_{\rm{ESS}}^k$ the binary variable indicating the charging operation and $0<\eta_{\rm{ESS}}^c,\eta_{\rm{ESS}}^d,\eta_{\rm{loss}}<1$. There exist an upper bound ($\overline{P}_{\rm{ESS}}$) and lower bound ($\underline{P}_{\rm{ESS}}^k$) on the $P_{\rm{ESS}}$ so that

\small
\begin{equation}\label{eq:PS}
-(1-\delta_{\rm{ESS}}^k)~\underline{P}_{\rm{ESS}} \le P_{\rm{ESS}}^k \le \delta_{\rm{ESS}}^k~\overline{P}_{\rm{ESS}}.
\end{equation}
\normalsize

\noindent In addition, an ESS cost has been considered  based on the charging-discharging cycle to understand the effect of storage degradation in scheduling MES \cite{sampath2017control},

\small
\begin{equation}
    C(P^k_{\rm{ESS}}) = \frac{c_{\rm{ESS}}}{2 n \times cap_{\rm{ESS}} } \left( \frac{\delta_{\rm{ESS}}^k P_{\rm{ESS}}^{+^k}}{\tau_{\rm{ESS}}^{^+}} - \frac{(1-\delta_{\rm{ESS}}^k) P_{\rm{ESS}}^{-^k}}{\tau_{\rm{ESS}}^{^-}} \right),
\end{equation}
\normalsize

\noindent where, $\tau_{\rm{ESS}}^{^{+}/-}$ is the average number of time instants for charging/discharging, $n$ is the average number of charging-discharging cycles before ESS reaches to its end of life, $cap_{\rm{ESS}}$ is the capacity and $c_{\rm{ESS}}$ is the purchase as well as installation cost of the ESS.

\subsection{Energy Exchanged with the Utility Grid}

The power exchanged with the utility grid $P_{\rm{GR}}^k$ is constrained by lower ($\underline{P}_{\rm{GR}}$) and upper bound ($\overline{P}_{\rm{GR}}$) on the consumption given by
\small
\begin{equation}\label{eq:PU}
-\underline{P}_{\rm{GR}} \le P_{\rm{GR}}^k \le \overline{P}_{\rm{GR}}.
\end{equation}
\normalsize
\noindent The cost of power exchanged with the utility grid is time-varying and and modelled as,
\small
\begin{equation}
C_{\rm{GR}}^k = C_{\rm{D}}^k + C_{\rm{RT}}^k,
\end{equation}
\normalsize
\noindent where $C_{\rm{D}}^k$  and $C_{\rm{RT}}^k$ denote the day-ahead and real-time demand-based price of energy respectively at time instant $k$.
\subsection{Renewable Energy Source (RES)}
The uncertain renewable generation due to the photo-voltaic panels is given by $P_{\rm{R}}^k$, which is directly fed into the model from the renewable energy sources present in the system.

\subsection{Load Balance Constraints}
The MES has to meet the electrical ($P_{\rm{L}}^k$) and thermal ($Q_{\rm{L}}^k$) loads; the balance constraints at a given time instant $k$ are
\small
\begin{subequations}\label{eq:LB}
\begin{align}
& P_{\rm{L}}^k = \sum_{i=1}^{n_g} \delta_{{\rm{G}},i}^k P_{{\rm{G}},i}^k+\delta_{{\rm{GT}}}^k P_{{\rm{GT}}}^k+P_{\rm{R}}^k+P_{\rm{GR}}^k- \delta_{\rm{ESS}}^k P_{\rm{ESS}}^{+^k}  &\nonumber\\  
&~~~~~~~~~~~~~~~~~~~~~~~~~-(1-\delta_{\rm{ESS}}^k) P_{\rm{ESS}}^{-^k} - P_{\rm{C}}^k - P_{\rm{AC}}^k,& \label{subeq14}\\
& Q_{\rm{L}}^k =\sum_{j=1}^{n_c} Q_{{\rm{C}},j}^k+ Q_{\rm{AC}}^k-(1-\delta_{\rm{TES}}^k)P_{\rm{TES}}^{-^k}-\delta_{\rm{TES}}^k P_{\rm{TES}}^{+^k} &\label{subeq15}
\end{align}
\end{subequations} 
\normalsize
\vspace{-6pt}
\subsection{Cost Function}
The objective is to reduce the energy costs and peak-demand by scheduling the different energy vectors so as to minimize

\small
\begin{align}\label{eq:Obj1}
&J^k= \sum_{i=1}^{n_g} C(P_{{\rm{G}},i}^k)+C(P_{\rm{GT}}^k)+C_{{\rm{GR}}}^k P_{{\rm{GR}}}^k + C(P^k_{\rm{ESS}}).
\end{align}
\normalsize 

\subsection{MPC Optimization Model}

Considering a prediction horizon $N_p$, the MES scheduling problem can be defined on the decision vector over the time horizon $\{k+1, \dots, k+N_p\}$ as $\mathbf{u}^{k,N_p} =[\mathbf{u}^{k+1},\dots, \mathbf{u}^{k+N_p}]$ with

\small
\begin{align}\label{eq:U}
&\mathbf{u}^{k}= [P_{{\rm{G}},i}^k, \dots , P_{{\rm{G}},n_g}^k, P_{\rm{GT}}^k,Q_{\rm{AC}}^k,\dot{m}_{{\rm{C}},1}^k, \dots, \dot{m}_{{\rm{C}},n_c}^k, P_{{\rm{ESS}}}^k,P_{{\rm{TES}}}^k, \nonumber\\
&~~~~P_{\rm{GR}}^k,\delta_{{\rm{G}},1}^k, \dots,\delta_{{\rm{G}},n_g}^k,\delta_{\rm{GT}}^k, \delta_{{\rm{ESS}}}^k, \delta_{{\rm{TES}}}^k,\gamma_{{\rm{C}},1}^k, \dots, \gamma_{{\rm{C}},n_c}^k].
 \end{align}
\normalsize

\noindent \small$\mathbf{s}^k = [T_{{\rm{TES}}}^k, {SoC}_{{\rm{ESS}}}^k, {{SoC}}_{{\rm{TES}}}^k]$\normalsize~ denotes the state variables of the system. The MPC optimization model for time-epoch $k$ is given by
\small
\begin{align}\label{eq:M}
&\mathcal{M}:~~~~~~\underset{\mathbf{u}^{k,N_p}}{\operatorname{min}}\;\; \sum_{k=k+1}^{k+N_p} J^k &  \nonumber\\
&\operatorname{s.t.}\;\;\text{constraints}~\eqref{G1},~\eqref{G2},~ \eqref{CG1},~\eqref{CG2},~\eqref{eq:ACOP},~\eqref{CG4},~\eqref{eq:chiller1}-\eqref{eq:chiller2},~\eqref{eq:TES},&\nonumber\\
&~~~~~~~~~~~~~~~~~~~~ \eqref{TES1},~\eqref{eq:PTlast},~\eqref{eq:PU},~\eqref{subeq13}, \eqref{eq:PS}, ~\eqref{eq:LB}; &\nonumber\\
&~~~~~~~\text{Binary constraints}~ \delta_{{\rm{G}},i}^k,\delta_{\rm{GT}}^k, \delta_{\rm{ESS}}^k, \delta_{\rm{TES}}^k,\gamma_{{\rm{C}},j}^k \in \mathbb{B},&\nonumber \\
&~~~~~~~~~~~~~~~~~~~~~~~~\forall k \in \{k+1,.., k+N_p\}. &
\end{align}
\normalsize
\noindent Eqn. \eqref{eq:M} implementing the MPC aims to obtain a control approach to perform PBDR by scheduling MES devices with fluctuating renewable generation, demand and energy prices. The MINLP-MPC optimization problem $\mathcal{M}$ possesses non-convexity, which needs to be solved over a prediction horizon with ramp rate constraints of the generation units \eqref{subeq4} and \eqref{subeq7} and time-coupled bilinear storage dynamics \eqref{eq:TES}. Thus, solving the problem directly with existing MINLP solvers such as BARON, Counene and other meta-heuristic algorithms is computationally intensive, which makes their adoption in a real-time EMS challenging \cite{burer2012,liberti2011recipe}. Also, heuristic solvers are only suitable for off-line computations due to their slow convergence.

\section{The Proposed SB3 Algorithm}
\label{sec:SB3}
\begin{figure}[h]
\centering
\includegraphics[scale= 0.28]{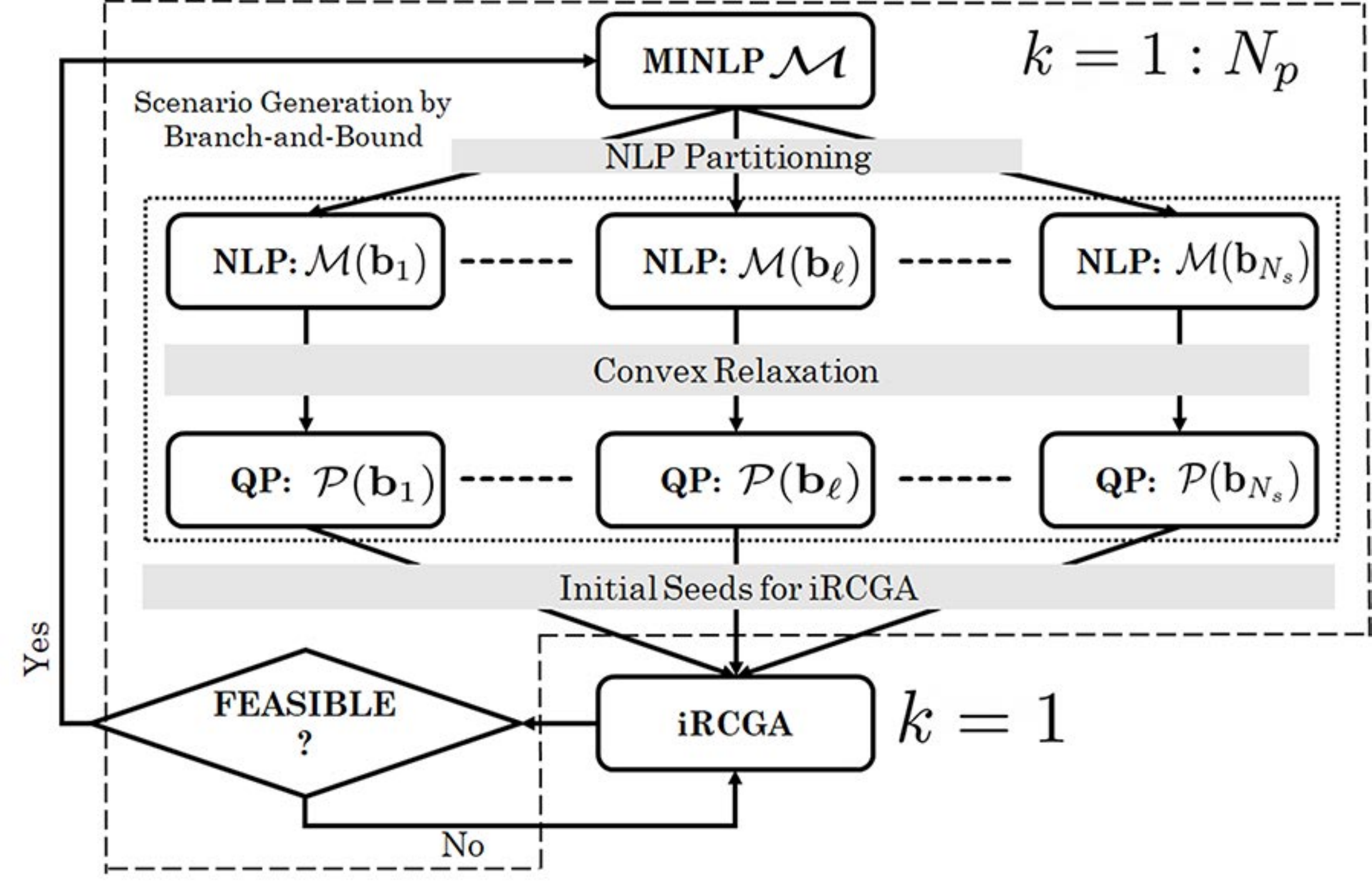}
\caption{Schematic diagram of SB3 algorithm}
\label{fig:SB3}
\end{figure}

The proposed light-weight solver, Scenario-based Branch-and-Bound (SB3), illustrated in Fig. \ref{fig:SB3} aims to address the aforementioned challenges. Widely used approaches to solve the MINLP are the generalized Benders decomposition, the outer approximation, the generalized outer approximation, and the extended cutting plane approach~\cite{lehmann2013efficient}. These methods linearize the MINLP by pivoting the binary constraints that leads to a sequential quadratic programming (SQP) or MILP formulation. In contrary, the proposed algorithm is accumulation of the following idea: 
\begin{enumerate}
	\item [(i)] to fix the valid binary decision variables to prior values with a priori knowledge and provide a set of scenarios by partially partitioning into a set of NLP sub-problems, followed by convex relaxation of the NLP sub-problems into a set of quadratic programming (QP) sub-problems over a prediction horizon as shown in Fig. \ref{fig:SB3},
	\item [(ii)] to use the resulting schedules from the scenario block as initial seeds to hot-start the MINLP solver, an improved real-coded genetic algorithm (iRCGA) to obtain scheduling for the first-time instant.
\end{enumerate}
The approach leads to a hierarchical problem with less computational complexity. The proposed algorithm is proposed with a motivation to find reasonable acceptable sub-optimal solution with a lower computational time. This section describes the steps of the SB3 algorithm.

\subsection{STEP 1: Scenario Generation and Branch-and-Bound - Seeds for MINLP Solver} \label{sec:seeds}
The importance of this step is to provide initial seeds to hot-start the iRCGA MINLP solver, which is intrinsically a population-based direct-search algorithm. The master problem $\mathcal{M}$ is divided into a set of convex QP problems by performing two operations sequenttially: {\em{(i)}} NLP partitioning and {\em{(ii)}} convex relaxation as discussed below.

\noindent \subsubsection*{(i) NLP Partitioning} 
Partitioning is an efficient technique for handling the combinatorial complexity of the problem. We define the following:

\noindent{\em{Definition~1:}} A partial partition of the $\mathcal{M}$ is a set of $N_s$ NLP sub-problems with $N_s \le N_q$ for which the definition of partition holds.

\noindent{\em{Definition~2:}} A valid binary string ${b}_\ell$ from the partial partition is the binary decision with $l_0$-norm $\|{b}_\ell\|_0 \leq n_b$, for which the combinatorial constraints in $\mathcal{M}$ are satisfied. $n_b$ is the total number of binary decision variables in $\mathcal{M}$.

\noindent \subsubsection*{{(ii) Convex Relaxation:}} 
The main challenge in the MINLP problem is the nonlinear constraints that needs to be relaxed without oversimplifying the analysis. The most restrictive constraints are the bilinear one in eqn. \eqref{eq:TES} and ramp rate constraints in eqn. \eqref{subeq4} and \eqref{subeq7}. Usual approaches such as SQP or a MILP formulation based on big-M method have been used~\cite{Verrilli2017}. We use a polyhedral over approximation to simplify the bilinear dynamics as a box constraint. In order to replace the bilinear term, we consider an auxiliary variable \small $\xi_j^k$\normalsize~such that \small $\gamma_{{\rm{C}},j}^k=1 \Leftrightarrow \xi_j^k = \dot{m}_{{\rm{C}},j}^k~T_{{\rm{TES}}}^k $\normalsize. The modified TES dynamics are given by:

\small
\begin{equation}
T_{{\rm{TES}}}^{k+1}=-\sum_{j=1}^{N_c}\mu \left( \frac{\xi_j^k}{V \rho}+ \frac{ \dot{m}_{{\rm{C}},j}^k T_{\rm{in}}^k}{V \rho} \right)+T_{\rm{TES}}^k(1-\frac{UA}{V \rho C_{\rho}})+\phi
\label{eq:TESdyn}
\end{equation}
\normalsize

\noindent where \small $\gamma_{{\rm{C}},j}^k=1 \Leftrightarrow \xi_j^k = \dot{m}_{{\rm{C}},j}^k~T_{{\rm{TES}}}^k $\normalsize and \small$\phi = T_{\rm{amb}}\left(\frac{UA}{V \rho C_{\rho}}\right)$\normalsize. Then the following constraints, also known as McCormick envelopes defined on \small $\xi_j^k$\normalsize~turns the problem with a bilinear term into a convex relaxed one:

\small
\begin{subequations}\label{eq:NeCons}
\begin{align}
&\xi_j^k \ge  \underline{\dot{m}}_{{\rm{C}},j}^k T_{{\rm{TES}}}^k + \underline{T}_{{\rm{TES}}}^k \dot{m}_{{\rm{C}},j}^k - \underline{T}_{{\rm{TES}}}^k \underline{\dot{m}}_{{\rm{C}},j}^k, &\\
&\xi_j^k \le \underline{\dot{m}}_{{\rm{C}},j}^k T_{{\rm{TES}}}^k+ \overline{T}_{{\rm{TES}}}^k \dot{m}_{{\rm{C}},j}^k - \overline{T}_{{\rm{TES}}}^k \underline{\dot{m}}_{{\rm{C}},j}^k, &\\
&\xi_j^k \ge \overline{\dot{m}}_{{\rm{C}},j}^k T_{{\rm{TES}}}^k+ \overline{T}_{{\rm{TES}}}^k \dot{m}_{{\rm{C}},j}^k - \overline{T}_{{\rm{TES}}}^k \overline{\dot{m}}_{{\rm{C}},j}^k, & \\
&\xi_j^k \le \overline{\dot{m}}_{{\rm{C}},j}^k T_{{\rm{TES}}}^k + \underline{T}_{{\rm{TES}}}^k \dot{m}_{{\rm{C}},j}^k - \underline{T}_{{\rm{TES}}}^k~ \overline{\dot{m}}_{{\rm{C}},j}^k. & 
\end{align}
\end{subequations}
\normalsize

\noindent McCormick envelopes provide a computationally fast convex relaxation of the bilinear terms compared to other existing methods. Similarly, using the epigraph technique, the ramp rate constraints are modelled as:

\small
\begin{subequations} \label{ramprel}
\begin{align}
&P_{{\rm{G}},i}^{k}-P_{{\rm{G}},i}^{k-1} - R_{{\rm{G}},i} \leq {\varepsilon}_{{\rm{G}},i}^{k},&\label{eq:ep1}\\
&P_{{\rm{G}},i}^{k}-P_{{\rm{G}},i}^{k-1} - R_{{\rm{G}},i} \geq {\varepsilon}_{{\rm{G}},i}^{k},&\label{eq:ep2}\\
&P_{\rm{GT}}^{k}-P_{{\rm{GT}}}^{k-1} - R_{{\rm{GT}}} \leq {\varepsilon}_{{\rm{GT}}}^{k},&\label{eq:ep3}\\
&P_{\rm{GT}}^{k}-P_{\rm{GT}}^{k-1} - R_{\rm{GT}} \geq {\varepsilon}_{\rm{GT}}^{k}.&\label{eq:ep4}
\end{align}
\end{subequations}
\normalsize

\noindent where, \small${\varepsilon}_{{\rm{G}},i}^{k},{\varepsilon}_{{\rm{GT}}}^{k} \ge 0 $\normalsize.~The introduction of the constraints \eqref{eq:NeCons}, linear TES dynamics \eqref{eq:TESdyn} and \eqref{ramprel} makes the non-convex NLP to have only linear constraints.

\noindent {\em{Definition~3:}} A {\it{Scenario}} $\mathcal{P}(\mathbf{b}_{\ell})$ is convex relaxation of the NLP obtained by pivoting the binary variables to a valid string ${b}_\ell$  followed by the relaxation of the non-linear constraints using Eqns.  \eqref{eq:NeCons} and \eqref{ramprel}. 

The set of convex QP sub-problems by pivoting the binary decision variable to a valid binary string $\mathbf{b}_\ell \in \mathbb{B}^{n_b}$ and convex relaxation is given by,

\small
\begin{align}
&\mathcal{P}(\mathbf{b}_{\ell}):\underset{\mathbf{u}_{{b}_{\ell}}^{k,N_p}}{\operatorname{min}}\;\; \sum_{k=k+1}^{k+N_p} \left( J^k +\lambda_1\sum_{i=1}^{n_g}{\varepsilon}_{{\rm{G}},i}^{k}+\lambda_2{\varepsilon}_{{\rm{GT}}}^{k} \right)&\label{eq:scenario}\\
&\operatorname{s.t.}~\text{constraints}~\eqref{G1},\eqref{subeq3},\eqref{eq:ep1},\eqref{eq:ep2},~ \eqref{CG1},~\eqref{eq:ep3},~\eqref{eq:ep4},~\eqref{subeq8},~\eqref{eq:ACOP},&\nonumber\\
&\eqref{CG4},~\eqref{eq:chiller1}-\eqref{eq:chiller2},~\eqref{eq:LB},~\eqref{TES1},~\eqref{eq:PTlast},~\eqref{eq:TESdyn},~\eqref{eq:NeCons},~\eqref{subeq13},~\eqref{eq:PS},~\eqref{eq:PU},&\nonumber\\
&\text{Binary constraints}~\delta_{{\rm{G}},i}^k,\delta_{\rm{GT}}^k, \delta_{\rm{ESS}}^k, \delta_{\rm{TES}}^k,\gamma_{{\rm{C}},j}^k \in \mathbf{b}_\ell & \nonumber\\
&~~~~~~~~~~~~~~~~~~~~~~~~\forall k \in \{k+1, \dots, k+N_p\}& \nonumber
\end{align} 
\normalsize
where, $\lambda_1$ and $\lambda_2$ are penalty factors. Epigraph terms in objective function in $\mathcal{P}(\mathbf{b}_{\ell})$ of eqn. \eqref{eq:scenario} takes care of the error introduced due to convex relaxation of the ramp-rate constraints. The problem $\mathcal{P}(\mathbf{b}_{\ell})$ of eqn. \eqref{eq:scenario} can be solved by general QP solvers efficiently to find initial seeds for iRCGA solver. 

\subsubsection{Branch-and-Bound (B2) for Scenario Generation} \label{sec:SB3B2}
The binary decision variables that ensures feasibility plus optimality from step \ref{sb3step2} of previous instant's scheduling problem is given as the input to the B2 step. The B2 expands each node of the tree by pivoting the binary variables considering only the temporal constraints. The purpose of expanding the branches of the tree based on the binary temporal variables is to provide a feasible scenario over the prediction horizon. At the leaf nodes, the convex MPC problems of eqn. \eqref{eq:scenario} generated after pivoting and relaxation are solved. The B2 procedure in association with QP keeps track of feasibility of the scenarios in terms of temporal, operating, physical as well as equality constraints for the current prediction horizon. A scenario is considered to be infeasible if there exist no solution of the QP problem at the leaf nodes and subsequently the corresponding scenario is pruned. The procedure repeats until a user-defined maximum number of feasible scenarios are obtained. The solution of MPC from each unpruned branch acts as initial seed to hot-start MINLP solver which solves $\mathcal{M}$ only for the current time-instant. The following example is used to illustrate the working principle of the aforementioned B2 process.
\begin{figure}[h]
	\centering
	\includegraphics[scale= 0.28]{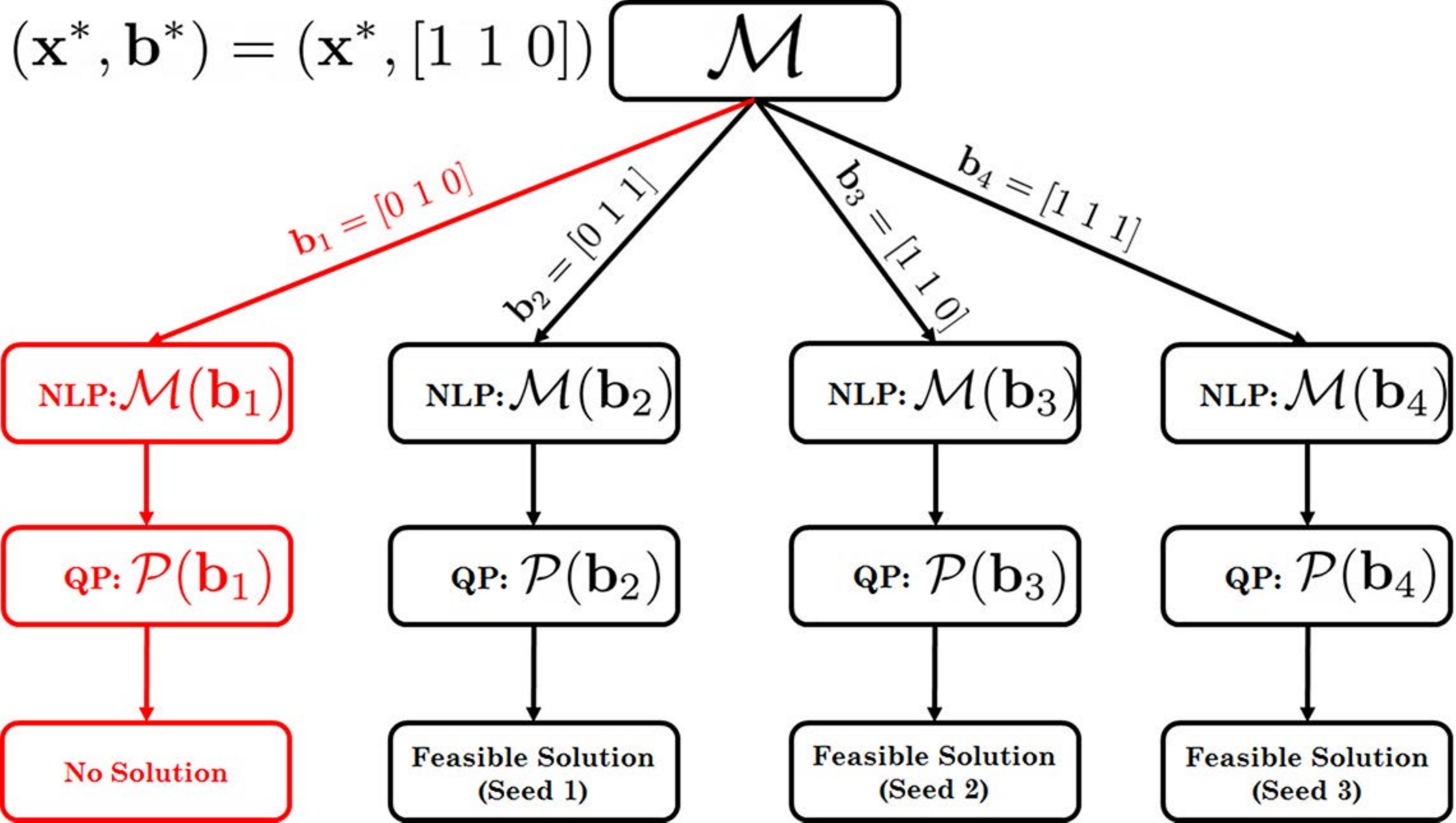}
	\caption{Illustration of B2 step in SB3 algorithm for scenario generation}
	\label{fig:B2illus}
\end{figure}

\noindent {{\bf{Illustrative example:}}} Let us consider a 2 DG and 1 GT system with $N_p = 1$ in which, $T^{\rm{up}}_{{\rm{G}},1}=2,~T^{\rm{up}}_{{\rm{G}},2}=2,~T^{\rm{up}}_{{\rm{GT}}}=2$ and $T^{\rm{down}}_{{\rm{G}},1}=2,~T^{\rm{down}}_{{\rm{G}},2}=2,~T^{\rm{down}}_{{\rm{GT}}}=2$. The solution from the previous instant is considered as $(\mathbf{x}^*,\mathbf{b}^*)$, where $\mathbf{b}^* = [1~1~0]$ with recorded consecutive ON-OFF time as $\tau^{\rm{up}}_{{\rm{G}},1}=2,~\tau^{\rm{up}}_{{\rm{G}},2}=1,~\tau^{\rm{down}}_{{\rm{GT}}}=1$. In Fig,~\ref{fig:B2illus}, the branches define the feasibility of the binary variables subject to the temporal constraints of the system only. However the existence of the solution of the QP problems after convex relaxation decides the feasible scenario. The branch associated with the binary variables $\mathbf{b}_1$ is pruned as it does not converge to a feasible solution. The solutions associated with the branches $\mathbf{b}_2,\mathbf{b}_3$ and $\mathbf{b}_4$ act as initial seeds for the MINLP solver.

\subsection{STEP 2: Solution of the MINLP solver with Feasibility Verification} \label{sb3step2} To solve the MINLP, we propose a novel improved real-coded genetic algorithm (iRCGA), a meta-heuristic optimization method which extends the capability of RCGA by providing adaptive selection for the genetic operators-- crossover and mutation (e.g., simple, heuristic, and arithmetic crossover etc., and binary mutation, Gaussian mutation etc. respectively) (Ref. supplementary material for detail discussion). As the number of feasible scenarios are large in number, solving the problem using mathematical optimization techniques for each of the initial seed is computationally intensive. In these circumstances, population based direct searching ability of meta-heuristic techniques from a small set of initial seeds are more efficient to find the solution of the MINLP problem. 

RCGA over GA is selected as it does not need phenotype to genotype conversion and hence it converges faster. Due to natural representation, one can achieve high precision and easy to integrate constraints in string representation. String length used in RCGA is small and computationally faster \cite{Dan2018ssci,Suresh2014}. The availability of multiple initial seeds provide iRCGA to search exhaustively through the search-space. Typically, in RCGA, different crossover and mutation techniques are widely used for generating off-spring vectors and for preserving diversity. Practice shows that selection of the crossover and mutation is problem dependent \cite{Suresh2014}. Therefore, in order to enhance the performance of conventional RCGA, we use an adaptive selection technique that increases RCGA's capability to tackle both continuous and integer variables simultaneously, without any relaxation. The selection of crossover and mutation techniques on basis of objective function value guides the algorithm towards faster convergence than using a particular crossover and mutation. We call the RCGA with multiple genetic operators selected using an adaptive selection technique as iRCGA. The detail of iRCGA algorithm can be found in the supplementary material for readers' understanding.

iRCGA is a population based direct search MINLP solver. In our approach, iRCGA solves a single instant MINLP instead of a multi-time step MPC problem. The solution of iRCGA, thus found, can lead to binary variables of incompatible temporal constraints over the prediction horizon considered earlier in section \ref{sec:seeds}. The feasibility verification step of the proposed SB3 algorithm ensures that the solution found from iRCGA has temporal constraint compatibility over the prediction horizon. Although feasibility verification follows similar approach of section \ref{sec:seeds}, it tries to check existence of a single scenario, which produces a feasible QP solution over the prediction horizon instead of finding a set of scenarios. It is to be noted that it only verifies the existence of feasible binary variables and the existence of such scenario does not necessarily guarantee the existence of MINLP solution over the prediction horizon. If temporal constraint feasibility is achieved, the solution from iRCGA is implemented at that time-instant. In case of failure, the iRCGA is iterated again for an alternative solution.

\subsection{Lower Bound Estimation of the SB3 Algorithm}
This two stage decision problem requires QP to provide good initial conditions for the MINLP solver. The following theorem provides the conditions under which the QP is able to produce lower bound solution (LBS) to $\mathcal{M}$.

\noindent {{\bf{Notation:}}} We define single instances of $\mathcal{M}$ and $\mathcal{P}(\mathbf{b}_\ell)$ at time instant $k$ as $\mathcal{M}^k$ and $\mathcal{P}^k$. To simplify our analysis, we use the following notations: $f(\cdot)$ denotes the objective function, the set of inequality constraints of $\mathcal{M}^k$ are denoted by the function $\mathbf{g}^{\mathcal{M}}$ and those of $\mathcal{P}^k$ by $\mathbf{g}^{\mathcal{P}}$. The real decision variables are denoted by $\mathbf{x} \in \mathbb{X} \subseteq \mathbb{R}^{m_c}$ with ${m_c}$ is the number of real variables and  the binary variables by $\mathbf{b} \in\mathbb{B}^{n_b}$. Without loss of generality, the equality constraints are replaced with corresponding appropriate inequality constraints for ease of the analysis.

\begin{Theorem} \label{lem1}
Suppose \small$\eta^{\mathcal{M}^*}$\normalsize ~ is the optimal solution of \small$\mathcal{M}^k$\normalsize ~ and \small$\eta^{\mathcal{P}^*}$\normalsize  is the LBS for a given set of scenarios \small$N_s$\normalsize ~ with  \small$\mathcal{P}^k(\mathbf{b}^*)$\normalsize ~ being the corresponding scenario. Then \small$\eta^{\mathcal{P}^*}$\normalsize is a LBS to \small$\mathcal{M}^k$\normalsize ~ if \small$\mathcal{M}^k$\normalsize ~ allows a convex continuous SQP relaxation with optimal solution with \small$\left( \mathbf{x}^{\mathcal{P}^*},\mathbf{b}^{\mathcal{P}^*} \right)$\normalsize ~ and there exists \small$\varOmega \geq 0$\normalsize ~ such that the following condition holds:

\small
	\begin{gather*}
	\varOmega \geq \operatorname{max}\{ M_B \tilde{\varOmega}, (1+M_{\nabla g})\tilde{\varOmega},  (1+ M_{\nabla g}) M_\lambda\tilde{\varOmega} \}, \nonumber
	\end{gather*}
	\normalsize
	
\noindent where \small$\|B\|_2 \leq M_B$, $ \|\nabla g_j^{\mathcal{P}}(X^{\mathcal{P}^*}, b^{\mathcal{P}^*})\|_2 \le M_{\nabla g}$, $|\lambda_j^\mathcal{P}| \leq M_\lambda,~\forall j \in \{1,...n_c\}$\normalsize .
\end{Theorem}

\noindent {\emph{Proof:}} 
Let $\eta^{\mathcal{P}^*}$ be the LBS obtained by solving the dual of $\mathcal{P}^k(\mathbf{b}_{\ell})$ computed as,
\begin{equation} \label{eq:QP}
\eta^{\mathcal{P}^*} \le   \underset{\ell \in N_s}{\operatorname{inf}} \left( \underset{\mathbf{\lambda} }{\operatorname{max}}  \left( \underset{\mathbf{x} \in \mathcal{X}}{\operatorname{min}} \left(f(\mathbf{x},\mathbf{b}_\ell)+ \mathbf{\lambda}^{\rm{T}} {\mathbf{g}^{\mathcal{P}}(\mathbf{x},\mathbf{b}_\ell)}\right)\right)\right)
\end{equation}
\noindent and $[\mathbf{x}^*$, $\mathbf{b}^*]$ and ${\mathbf{\lambda}}^*$ denote the corresponding primal solution and Lagrange multipliers of the inequality and equality constraints, respectively. The Lagrangian of $\mathcal{M}^k$ is given by

\small
\begin{align} \label{eq:NLP1}
&\mathcal{L}^{\mathcal{M}} = f(\mathbf{x},\mathbf{b})+ {\mathbf{\lambda}}^{\rm{T}} {\mathbf{g}^{\mathcal{M}}(\mathbf{x},\mathbf{b})}.
\end{align}
\normalsize

\noindent Let us consider that $\mathbf{\tilde{x}}$ denotes the solution obtained by fixing the binary variable to $\mathbf{b}_\ell$ in \eqref{eq:NLP1} from the scenario set $\{ \mathbf{b}_\ell, \ell = 1,\dots,N_s \}$ and the corresponding solution is an upper bound to the problem $\mathcal{M}^k$ with objective value as $\tilde{\eta}$ \cite{lehmann2013efficient}. Thus,

\small
\begin{equation} \label{eq:QP}
\eta^{\mathcal{M}^*} \le   \underset{\ell \in N_s}{\operatorname{inf}} \left( \underset{\mathbf{\lambda} }{\operatorname{max}}  \left( \underset{\mathbf{x} \in \mathcal{X}}{\operatorname{min}} \left(f(\mathbf{x},\mathbf{b}_\ell)+ \mathbf{\lambda}^{\rm{T}} {\mathbf{g}^{\mathcal{M}}(\mathbf{x},\mathbf{b}_\ell)}\right)\right)\right) \le \tilde{\eta},
\end{equation}
\normalsize

\noindent where, $\eta^{\mathcal{M}^*}$ is the optimal solution of $\mathcal{M}^k$. Defining $\mathbf{d} =[\mathbf{d_c}~|~\mathbf{d_b}]= [\mathbf{x-\tilde{x}}~|~ \mathbf{b- {b}_\ell}]$, the trust-region continuous SQP approximation \cite{exler2012} of $\mathcal{M}_k$ around $[ \mathbf{\tilde{x}}, \mathbf{b}_\ell]$ with convex relaxation is given by,

 \small
 \begin{align}\label{eq:SQP}
 & \underset{\mathbf{d},\nu^{\mathcal{P}}} {\operatorname{min}} ~~~~\nabla f(\mathbf{\tilde{x}},\mathbf{b}_\ell)^{\rm{T}} \mathbf{d}+\frac{1}{2} \mathbf{d}^{\rm{T}} {\rm{B}} \mathbf{d} +\sigma \nu^{\mathcal{P}} \nonumber\\
 & \operatorname{s.t.}~~~\mathbf{g}^{\mathcal{P}}(\mathbf{\tilde{x}},\mathbf{b}_\ell)+\nabla \mathbf{g}^{\mathcal{P}}(\mathbf{\tilde{x}},\mathbf{b}_\ell)^{\rm{T}} \mathbf{d} \leq \nu^{\mathcal{P}} \mathbf{1} \\
 &~~~~~~~ \|\mathbf{d}\|_\infty \leq \Delta, ~\mathbf{x+d_c} \in \mathcal{X},~\mathbf{b+d_b} \in \{\mathbf{b}_{\ell}, \ell = 1,\dots,N_s\} \nonumber 
\end{align}
\normalsize

\noindent where $\sigma$ is a positive penalty term and $\nu^{\mathcal{P}} = \operatorname{max}\left(0,~\|\mathbf{g}^{\mathcal{P}}(\mathbf{\tilde{x}},\mathbf{{b}}_\ell)+\nabla \mathbf{g}^{\mathcal{P}}(\mathbf{\tilde{x}},\mathbf{b}_\ell)^{\rm{T}} \mathbf{d}\|_\infty \right)$. ${\rm{B}}$ is the quassi-Newton approximaion of the hessian of the Lagrangian. The solution from \eqref{eq:SQP} provides a LBS to \eqref{eq:NLP1} due to convex continuous relaxation. The SQP formulation allows for handling the infeasible constraints including the temporal constraints as well as temporal search subspace. After a certain number of iterations, it provides a solution denoted by $\left( \mathbf{x}^{\mathcal{P}^*},\mathbf{b}^{\mathcal{P}^*} \right)$ which is closer to $(\mathbf{x}^*,\mathbf{b}^*)$ by construction~\cite{exler2012}. It is to be mentioned that pivoting the binary variable to a particular value of binary variables produces a convex continuous optimization problem of \eqref{eq:QP}. We define the following 

\begin{gather*}
\|\mathbf{x}^{{\mathcal{P}}^*}-\mathbf{x}^*\|_2 \leq \varOmega,~~ \nu^{\mathcal{P}} \leq \tilde{\varOmega}_{\nu},~~\|\mathbf{d}\|_2 \leq \tilde{\varOmega}_d.
\end{gather*}

Considering $\tilde{\varOmega} = \operatorname{max} \{ \tilde{\varOmega}_d, \tilde{\varOmega}_{\nu} \} $, from the necessary conditions of optimality for the KKT point $\left((\mathbf{x}^{\mathcal{P}^*},\mathbf{b}^{\mathcal{P}^*}),\mathbf{\lambda}^{\mathcal{P}^*}\right)$ with dual solution being $\mathbf{\lambda}^{\mathcal{P}^*}$, we have

\small
\begin{align}
&\nabla f(\mathbf{x}^{\mathcal{P}^*}, \mathbf{b}^{\mathcal{P}^*}) + {\rm{B}}~\mathbf{d} +\sum_j \lambda_j^{\mathcal{P}} \nabla g_j(\mathbf{x}^{\mathcal{P}^*}, \mathbf{b}^{\mathcal{P}^*})= 0, \nonumber \\
\implies&\|\nabla f(\mathbf{x}^{\mathcal{P}^*}, \mathbf{b}^{\mathcal{P}^*})+\sum_j \lambda_j^{\mathcal{P}} \nabla g_j(\mathbf{x}^{\mathcal{P}^*}, \mathbf{b}^{\mathcal{P}^*})\|_2 \leq \|{\rm{B}}\|_2 \|{\mathbf{d}}\|_2 \nonumber \\
&~~~~~~~~~~~~~~~~~~~~~~~~~~~~~~~~~~~~~~~~~~~~~~~~~\leq M_B \tilde{\varOmega}
\end{align}
\normalsize

\noindent where $\|{\rm{B}}\|_2 \leq M_B$ is the bound on the Hessian. Considering the primal feasibility condition,

\small
\begin{align}
& g_j^{\mathcal{P}}(\mathbf{x}^{\mathcal{P}^*}, \mathbf{b}^{\mathcal{P}^*})+  \nabla g_j^{\mathcal{P}}(\mathbf{x}^{\mathcal{P}^*}, \mathbf{b}^{\mathcal{P}^*})^{\rm{T}} \mathbf{d} \leq \nu^{\mathcal{P}} \nonumber \\
\implies& g_j^{\mathcal{P}}(\mathbf{x}^{\mathcal{P}^*}, \mathbf{b}^{\mathcal{P}^*}) \leq -\nabla g_j^{\mathcal{P}}(\mathbf{x}^{\mathcal{P}^*}, \mathbf{b}^{\mathcal{P}^*})^{\rm{T}} \mathbf{d}+\nu^\mathcal{P} \nonumber \\
\implies&\| g_j^{\mathcal{P}}(\mathbf{x}^{\mathcal{P}^*}, \mathbf{b}^{\mathcal{P}^*})\|_2\leq \|\nabla g_j^{\mathcal{P}}(\mathbf{x}^{\mathcal{P}^*}, \mathbf{b}^{\mathcal{P}^*})\|_2 \|\mathbf{d}\|_2+ \nu^\mathcal{P} \nonumber\\
&~~~~~~~~~~~~~~~~~~~~\leq {\tilde{\varOmega} (1+ M_{\nabla g})}
\end{align}
\normalsize

\noindent From dual feasibility condition, $\lambda_j^{\mathcal{P}} \ge 0, \forall j $ holds. Considering the complimentary slackness condition, we have

\small
\begin{align}
&\lambda_j^{\mathcal{P}}\left( g_j^{\mathcal{P}}(\mathbf{x}^{\mathcal{P}^*}, \mathbf{b}^{\mathcal{P}^*})+ \lambda_j^{\mathcal{P}} \nabla g_j^{\mathcal{P}}(\mathbf{x}^{\mathcal{P}^*}, \mathbf{b}^{\mathcal{P}^*})^{\rm{T}} \mathbf{d}\right) = \lambda_j^{\mathcal{P}}\nu^{\mathcal{P}} \nonumber \\
\implies& \lambda_j^{\mathcal{P}} g_j^{\mathcal{P}}(\mathbf{x}^{\mathcal{P}^*}, \mathbf{b}^{\mathcal{P}^*})= - \lambda_j^{\mathcal{P}} \nabla g_j^{\mathcal{P}}(\mathbf{x}^{\mathcal{P}^*}, \mathbf{b}^{\mathcal{P}^*})^{\rm{T}} \mathbf{d} + \lambda_j^{\mathcal{P}}\nu^{\mathcal{P}} \nonumber \\
\implies&  \|\lambda_j g_j^{\mathcal{P}}(\mathbf{x}^{\mathcal{P}^*}, \mathbf{b}^{\mathcal{P}^*})\|_2  \leq \|\lambda_j^{\mathcal{P}}\|_2 \left(\|\nu^{\mathcal{P}}\|_2+\|\nabla g_j^{\mathcal{P}}(\mathbf{x}^{\mathcal{P}^*}, \mathbf{b}^{\mathcal{P}^*})\|_2 \| \mathbf{d} \|_2 \right) \nonumber\\ 
&~~~~~~~~~~~~~~~~~~~~~~~\leq {(1+M_{\nabla g})M_\lambda \tilde{\varOmega}} 
\end{align}
\normalsize

\noindent Thus the SQP relaxation in \eqref{eq:SQP} produces a close proximal solution of \eqref{eq:QP} to an accuracy $\varOmega$ such that the following necessary conditions holds,

\begin{gather} \label{eq:nlpconclusion}
\varOmega \geq \operatorname{max}\{ M_B \tilde{\varOmega}, (1+M_{\nabla g}) \tilde{\varOmega},  (1+ M_{\nabla g}) M_\lambda\tilde{\varOmega} \}
\end{gather} 

\noindent with $\eta^{\mathcal{P}^*} \leq \eta^{\mathcal{M}^*} \leq \tilde{\eta}$. \eqref{eq:nlpconclusion} provides a lower bound accuracy measurement to the optimality of the problem. The low value of $\varOmega$ indicates the possibility of improving optimality of the solution.
\hfill $\bDiamond$

\noindent The Corollary 1 relates KKT points of SQP and $\mathcal{P}^k$.

\begin{Corollary}
For sufficiently large value of the penalty factor $\sigma$ and for some $\mathbf{b}=\mathbf{{b}}_\ell$, the KKT points of $\mathcal{P}^k$ coincides with that of the SQP relaxed problem with $\mathbf{d=0}$ and $\nu^{\mathcal{P}}=0$, where $\mathbf{d}$ is distance measurement in SQP formulation and $\nu^{\mathcal{P}}$ is the penalty factor for constraint violation.
\end{Corollary}

\noindent {\it{Proof:}} The KKT conditions of $\mathcal{P}^k$ at the optimal solution point are given by

\small
\begin{align}\label{eq:KKTNP}
& \nabla f(\mathbf{x}^*,\mathbf{b}^*)+\sum_j \lambda_j^* \nabla g_j^{\mathcal{P}}(\mathbf{x}^*,\mathbf{b}^*) =0, ~~ g_j^{\mathcal{P}}(\mathbf{x}^*,\mathbf{b}^*) \leq 0   \nonumber \\
& \lambda_j^* \geq 0, ~~~~\lambda_j^* g_j^{\mathcal{P}}(\mathbf{x}^*,\mathbf{b}^*)  = 0, ~\forall j.
%&\nonumber \\
\end{align}
\normalsize

Similarly, the KKT conditions of the SQP after a finite number of iteration (${\rm{iter}}$) are given by

\small
\begin{align}\label{eq:KKTSQP}
& \nabla f(\mathbf{x}^{\mathcal{P}^{\rm{iter}}}, \mathbf{b}^{\mathcal{P}^{\rm{iter}}})+ {\rm{B}}^{\rm{iter}} \mathbf{d}^{\rm{iter}}  + \sum_j \lambda_j^{\mathcal{P}^{\rm{iter}}} \nabla g_j^{\mathcal{P}}(\mathbf{x}^{\mathcal{P}^{\rm{iter}}}, \mathbf{b}^{\mathcal{P}^{\rm{iter}}}) =0\nonumber \\
& \sigma^{\rm{iter}} = \sum_j \lambda_j^{\mathcal{P}^{\rm{iter}}} \nonumber \\ &g_j^{\mathcal{P}}(\mathbf{x}^{\mathcal{P}^{\rm{iter}}}, \mathbf{b}^{\mathcal{P}^{\rm{iter}}})+\nabla g_j^{\mathcal{P}}(\mathbf{x}^{\mathcal{P}^{\rm{iter}}}, \mathbf{b}^{\mathcal{P}^{\rm{iter}}})^{\rm{T}} \mathbf{d}^{\rm{iter}} \leq \nu^{\mathcal{P}^{\rm{iter}}}\mathbf{1}    \nonumber \\
& \lambda_j^{\mathcal{P}^{\rm{iter}}} \left( g(\mathbf{x}^{\mathcal{P}^{\rm{iter}}}, \mathbf{b}^{\mathcal{P}^{\rm{iter}}})+\nabla g(\mathbf{x}^{\mathcal{P}^{\rm{iter}}}, \mathbf{b}^{\mathcal{P}^{\rm{iter}}})^{\rm{T}} \mathbf{d}^{\rm{iter}} \right)  = 0 \nonumber \\
& \lambda_j^{\mathcal{P}^{\rm{iter}}} \geq 0,~~\forall j
\end{align}
\normalsize

\noindent Now using the conditions that $\mathbf{d}^{\rm{iter}} = \mathbf{0}$, $\nu^{\mathcal{P}^{\rm{iter}}} = 0$ and and $\sigma^{\rm{iter}} \in \mathbf{R}^+$  in \eqref{eq:KKTNP}, we can rewrite the KKT conditions as,

\small
\begin{align}\label{eq:KKTSQP1}
& \nabla f(\mathbf{x}^{\mathcal{P}^{\rm{iter}}}, \mathbf{b}^{\mathcal{P}^{\rm{iter}}})  + \sum_j \lambda_j^{\mathcal{P}^{\rm{iter}}} \nabla g_j^{\mathcal{P}}(\mathbf{x}^{\mathcal{P}^{\rm{iter}}}, \mathbf{b}^{\mathcal{P}^{\rm{iter}}}) =0\nonumber \\
& \sigma^{\rm{iter}} = \sum_j \lambda_j^{\mathcal{P}^{\rm{iter}}} ,~~g_j^{\mathcal{P}}(\mathbf{x}^{\mathcal{P}^{\rm{iter}}}, \mathbf{b}^{\mathcal{P}^{\rm{iter}}}) \leq 0    \nonumber \\
& \lambda_j^{\mathcal{P}^{\rm{iter}}} g(\mathbf{x}^{\mathcal{P}^{\rm{iter}}}, \mathbf{b}^{\mathcal{P}^{\rm{iter}}})  = 0,~~\lambda_j^{\mathcal{P}^{\rm{iter}}} \geq 0,~~\forall j
\end{align}
\normalsize

Thus using \eqref{eq:nlpconclusion}, for $\mathbf{d}^{\rm{iter}} = \mathbf{0}$, $\nu^{\mathcal{P}^{\rm{iter}}} = 0$ and and $\sigma^{\rm{iter}} \in \mathbf{R}^+$, $\varOmega \ge 0$. Thus, $\exists$ $\mathbf{b}^{\mathcal{P}^{\rm{iter}}} = \mathbf{b}^{\mathcal{P}^*}$, at which the KKT conditions of both QP and SQP approaches coincide.   \hfill $\bDiamond$

\section{Results} 
\label{Sec:Results}
\begin{table}[htb!]
\caption{Rating of different MES components}
\centering
{\renewcommand{\arraystretch}{1.3}
\begin{tabular}{l|l|l}
\hline
Component & Rating & Constraints\\
\hline 
PV Panel  & $165$ \si{\kilo\watt}  & 2 units and Non-dispatchable\\
Electrical Storage & 300 \si{\kilo\watt\hour}& SoC: 30-90\%, $P_s = [-30,200]$ \si{\kilo\watt}\\
Generator Units & $1.2 $ \si{\mega\watt} & 2 units, $\underline{P_g}=400, \overline{P_g}=1200$ \si{\kilo\watt}\\
Gas turbine   & 1 \si{\mega\watt} & $\underline{P}_{GT} =250$ - $\overline{P}_{GT} = 1000$ \si{\kilo\watt}\\
Thermal Vector & 5400 \rm{RT} & 5 chillers, CoP = [2.8-3.3]\\
Absorption chiller &  350 \rm{RT} & $\underline{Q}_{AC}=200, \overline{Q}_{AC} =350$ \rm{RT} \\
TES & 2000 \rm{RTh} & ${\rm{SoC}}_T = [100,7000]$ \si{\kilo\watt\hour}, \\
&&$P_T$=[-1000, 3000] \si{\kilo\watt}\\
\hline
\end{tabular}}
\label{Tab:MES Components}
\end{table}

\begin{figure} [h]
	\centering
	\includegraphics [scale = 0.25]{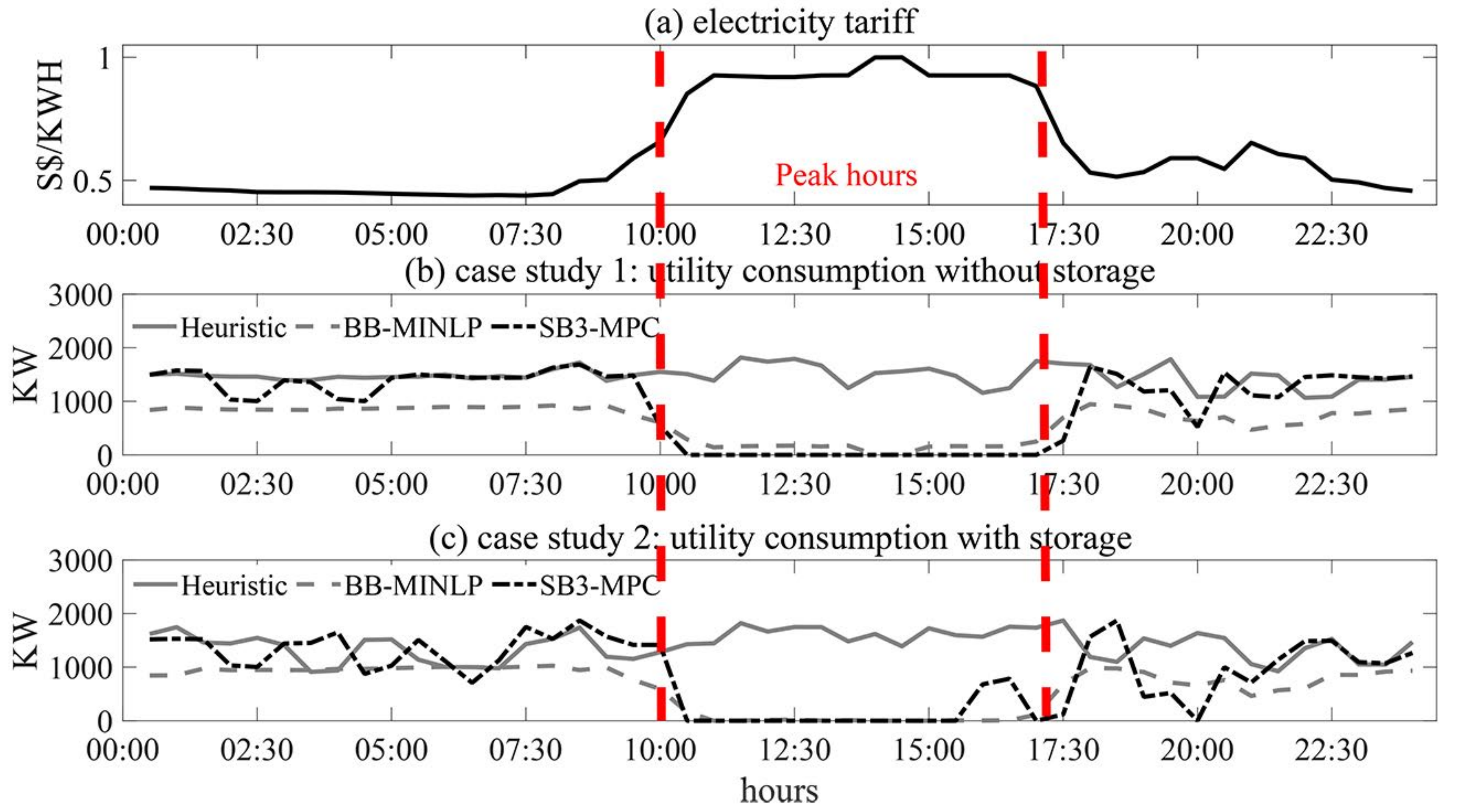}
	\caption{Demand response performance: SB3 v. heuristic-- (a) electricity tariff, (b) case study 1: without storage unit, (c) case study 2: with storage unit}
	\label{fig:DR}
\end{figure}

The PBDR performance of the proposed SB3 is illustrated on the CleanTech building, Singapore whose MES architecture is shown in Fig.~\ref{fig:MES}. The ratings and operating limits of the MES components are shown in Tab.~\ref{Tab:MES Components}. The following control strategies are compared: {\em{(i)}} heuristic optimization: The RCGA solves the MINLP problem $\mathcal{M}$ for multiple time-steps (prediction horizon) with receding horizon approach, {\em{(ii)}} branch-and-bound MINLP (BB-MINLP) with MPC, {\em{(iii)}} SB3 based MPC: it is the feedback control law computed in receding horizon manner solving the MINLP problem using the SB3 approach assuming accurate forecasts on renewable generation and demand, i.e., less than 2\% error, {\em{(iv)}} SB3 based MPC with forecast errors: it is the 24 hour feedback control law obtained by solving the MES scheduling problem using SB3 but with forecast errors of 2\% and 10\% to study the robustness, and {\em{(v)}} performance of SB3 based MPC without (case 1) and with (case 2) energy storage components. In addition, the sensitivity of the proposed SB3 algorithm is also analysed by varying the maximum number of scenarios during optimization.

In our study, a sampling time of 0.5 {\si{\hour}} with 10 {\si{\hour}} prediction horizon and 24 {\si{\hour}} control horizon is selected. Further, the renewable energy source is non-dispatchable and the peak-demand period is from 10 AM (10:00 {\si{\hour}}) to 5.00 PM (17:00 {\si{\hour}}) computed from daily load-curve as shown in supplementary material. The user-defined number of scenarios is kept at 20 for all the simulations to keep the simulation time tractable. The heuristic optimization is considered with multi-time step receding horizon approach however, its large computational time hinders its implementation in real-time systems \cite{morvaj2017,Dan2018ssci}.

\subsubsection*{\textbf{Note on forecasting}} Implementing MPC for scheduling MES components requires forecasts on renewable generation and demand. Meta-Cognitive Fuzzy Inference System (McFIS) is a neuro-fuzzy inference approach, which is based on an adaptive sequential learning algorithm that starts approximating a non-linear function with zero fuzzy rule and develops the necessary number of fuzzy rules depending upon the information contained in the training samples (historical forecasting data). A detailed analysis of the forecasting methodology using McFIS can be found in~\cite{dan2018}. 

\subsection{Demand Response}
The PBDR based scheduling of MES heuristic optimization (RCGA), BB-MINLP and the SB3 approach is studied. Fig.~\ref{fig:DR}~(a) shows the normalized energy price for 24 {\si{\hour}} which are high during peak-periods i.e., 10 AM (10:00 hr) to 5.00 PM (17:00 hr). It can be observed in Fig.~\ref{fig:DR}(b) and \ref{fig:DR}(c) as well as in Tab.~\ref{tab:DR} that during peak-hours the SB3 based MPC reduces energy utilization from utility grid by 82.37\% and 96.78\% in case study 1 and 2, respectively. Whereas, heuristic method reduces only 19.57\% and 10.46\%, respectively as a result of pre-mature termination and BB-MINLP reduces 76.06\% and 92.58\%, respectively. Thus SB3-MPC provides a promising improvement in demand response with respect to the heuristic algorithm as well as the BB-MINLP.

\begin{table}
\centering
\caption{Comparison of price-based demand response (PBDR) performance}
{\renewcommand{\arraystretch}{1.1}
\begin{tabular}{|c|cc|cc|}
\hline
 Control approach & without storage & with storage\\ \hline
Heuristic & 19.57 & 10.46  \\
BB-MINLP & 76.06 & 92.58 \\ 
SB3-MPC & 82.37 & 96.78 \\ \hline
\end{tabular}}
\label{tab:DR}
\end{table}
\begin{figure}
	\centering
	\includegraphics [scale = 0.2]{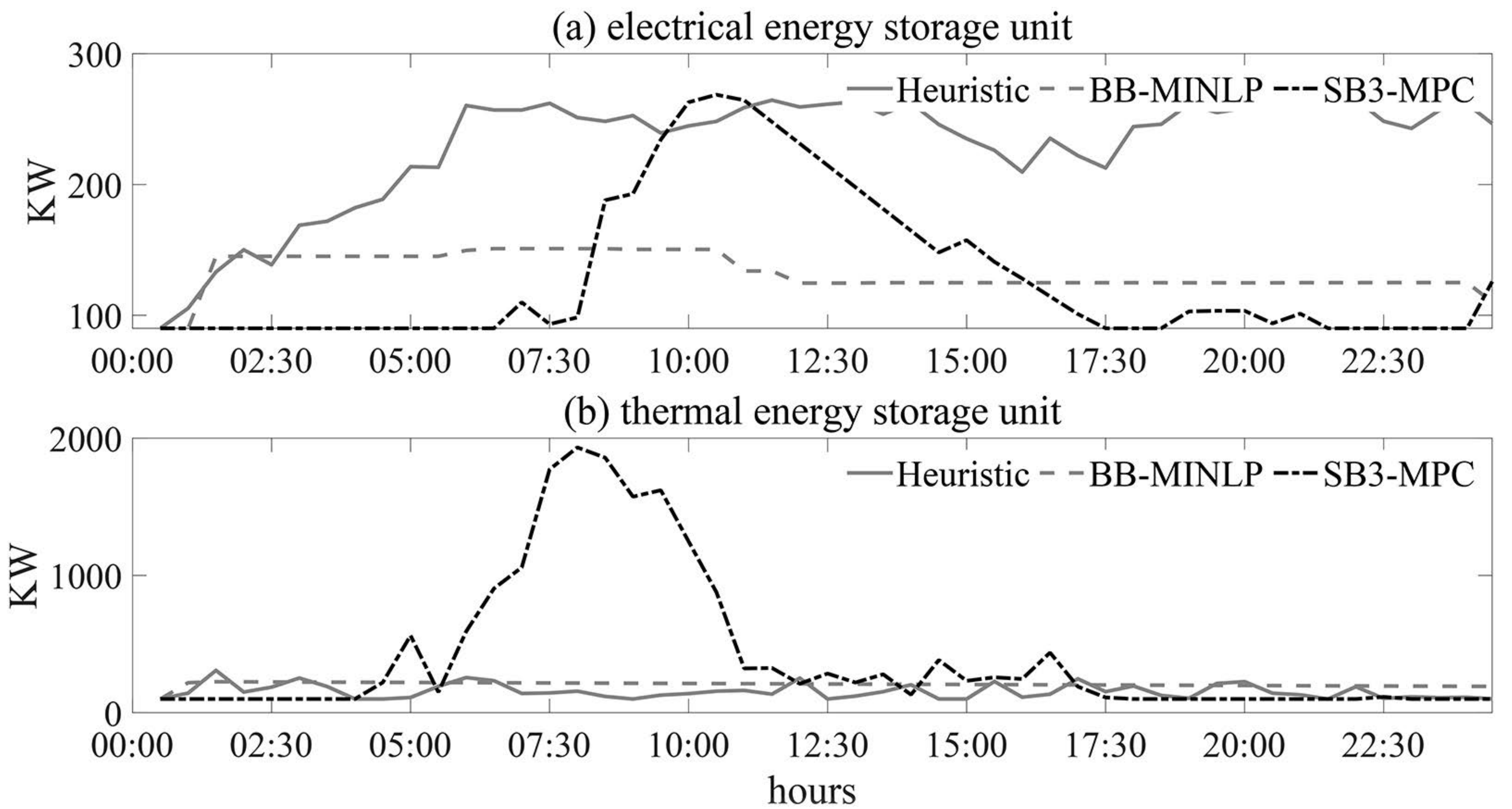}
	\caption{(a) ESS and (b) TES management}
	\label{fig:SOC}
\end{figure}
\begin{figure}
    \centering
    \includegraphics[scale = 0.2]{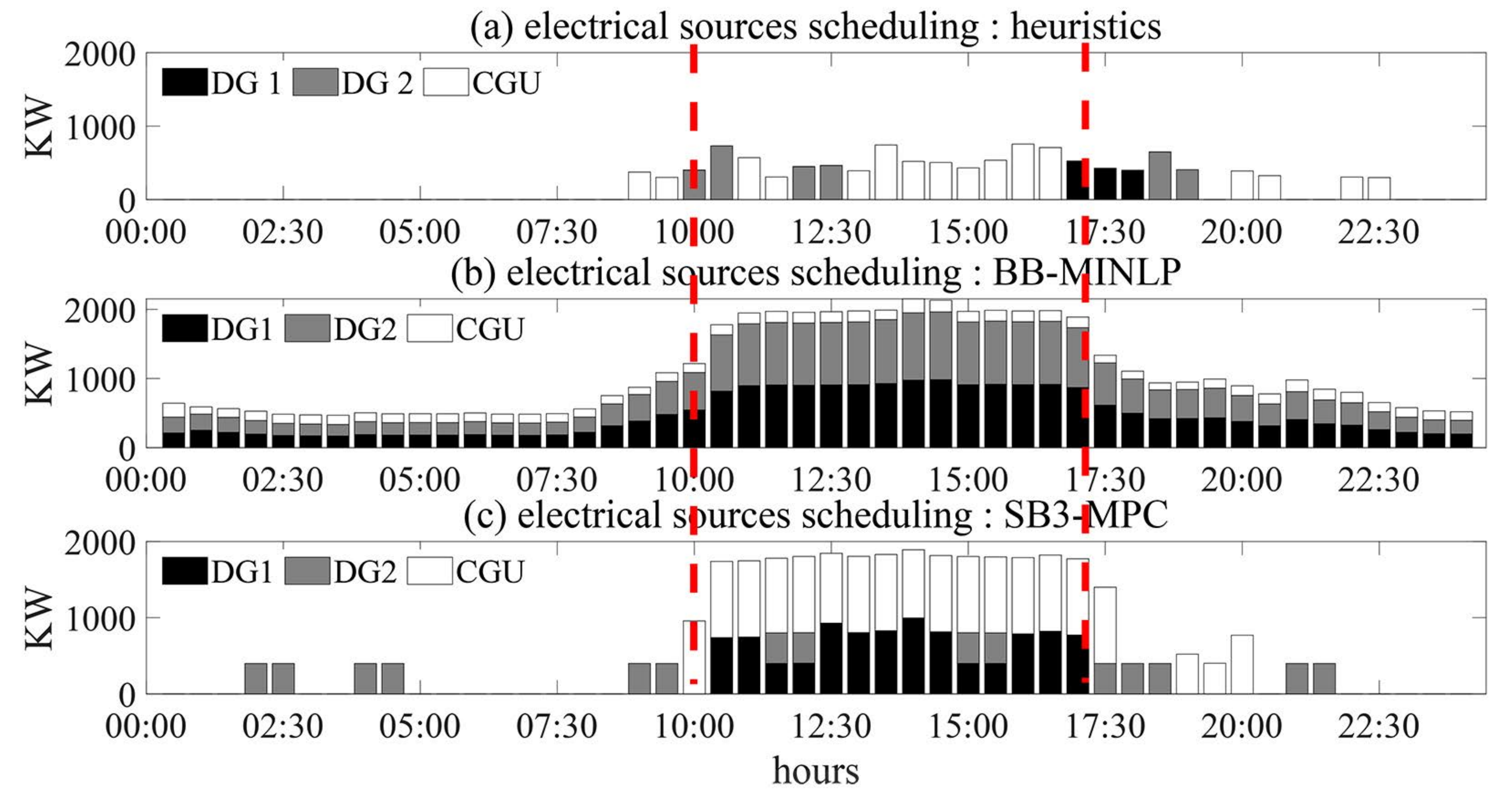}
        \caption{Electrical sources scheduling: case study 1 - without storage unit}
    \label{fig:ESwos}
\end{figure}

\subsection{Storage Management and Cost-efficient MES Scheduling}
Flexibility provided by storage devices can be used to augment capacity and deal with uncertain elements more effectively. The case-study considered has both electrical and thermal storage, the variations in SoC of the thermal and electrical storage are shown in Fig. \ref{fig:SOC}. It can be verified that the SB3 based MPC increases the SoC of the storage devices during off-peak load periods when the price is low and discharges it during peak-demand periods. As against this, the heuristic control and BB-MINLP both under utilize the storage units during peak-price hours. BB-MINLP restricts the use of the storage units due to presence of storage degradation costs. A poor demand response performance of heuristic algorithm as well as its incapability of scheduling the local generation units during peak-price hours have been caused due to its pre-mature simulation termination at maximum iteration.
\begin{figure}
    \centering
    \includegraphics[scale = 0.25]{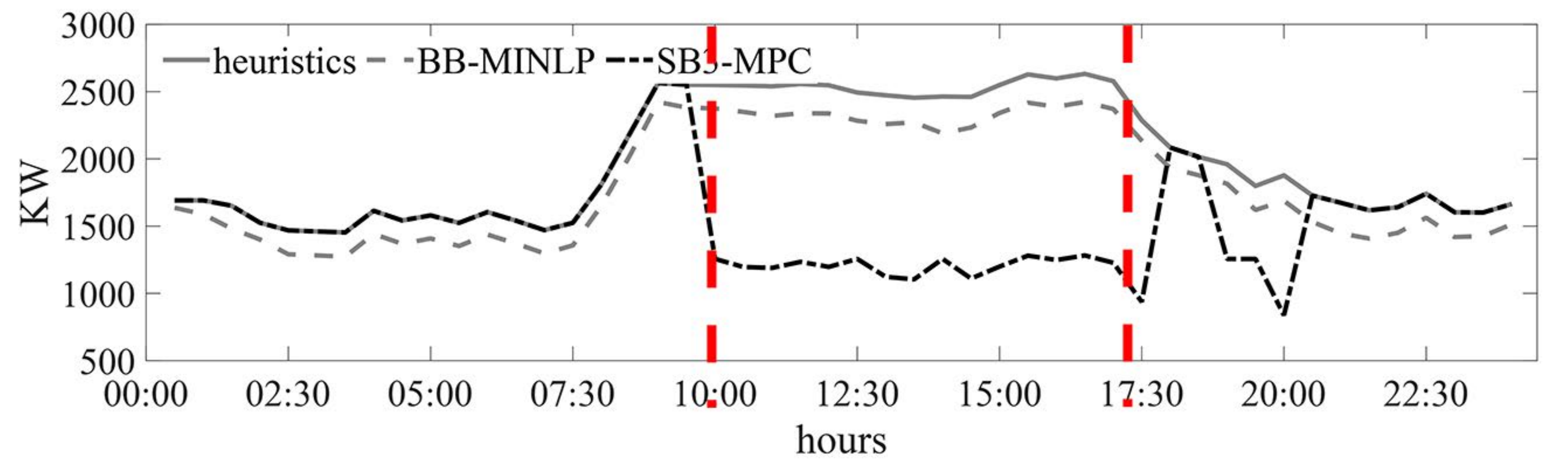}
        \caption{Chiller bank scheduling: case study 1 - without storage unit}
    \label{fig:cbwos}
\end{figure}
\begin{table}
	\centering
	\caption{Performance of SB3 with forecast errors (FE)}
	{\renewcommand{\arraystretch}{1}
	\begin{tabular}{|c|c|c|}
		\hline 
		Control approach& Change in demand-response & Change in cost\\
		\hline 
		SB3-MPC with 2\% FE & $\pm 0.08\%$ & $\rm{\pm 1.21 \%}$  \\
		SB3-MPC with 10 \% FE & $\pm 1.77\%$ & $\pm 2.34\% $\\ 
		\hline
	\end{tabular}}
	\label{tab:perf}
\end{table}

\begin{figure}
    \centering
    \includegraphics[scale = 0.2]{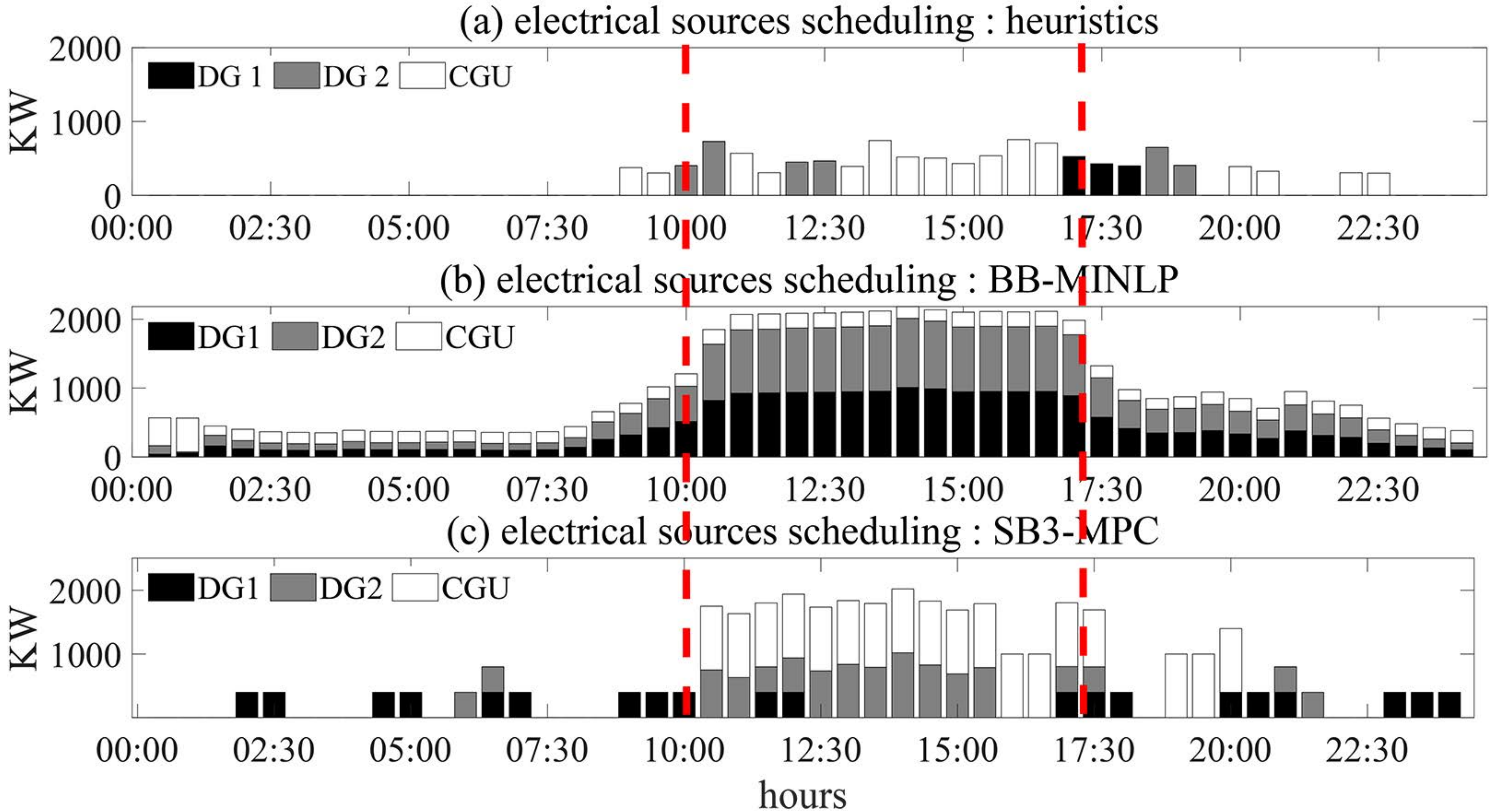}
        \caption{Electrical sources scheduling: case study 2 - with storage unit}
    \label{fig:ESws}
\end{figure}

\begin{figure}
    \centering
    \includegraphics[scale = 0.25]{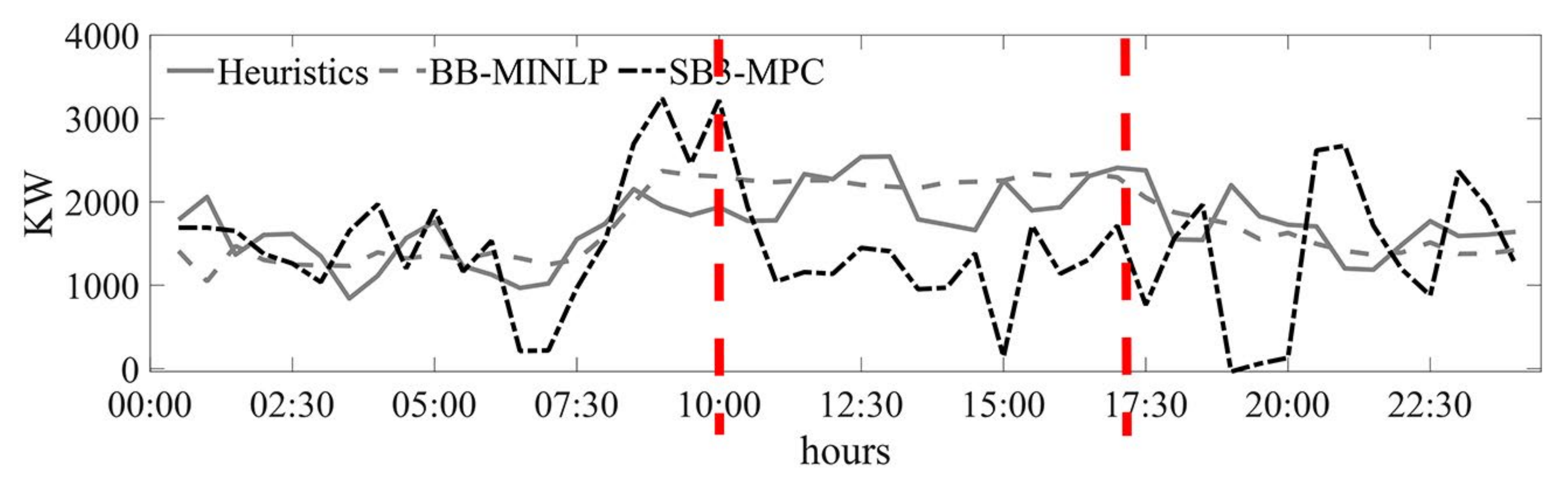}
        \caption{Chiller bank scheduling: case study 2 - with storage unit}
    \label{fig:cbws}
\end{figure}

The MES component schedules are shown in Fig.~\ref{fig:ESwos}, \ref{fig:cbwos} without storage units and in Fig.~\ref{fig:ESws}, \ref{fig:cbws} with storage units. The SB3-based MPC schedules local energy sources strategically depending on the price vector to minimize the cost. Further, local energy source utilization increased during peak-periods as shown in Fig.~\ref{fig:ESwos}(c) and Fig.~\ref{fig:ESws}(c) compared to heuristic algorithm in Fig.~\ref{fig:ESwos}(a) and Fig.~\ref{fig:ESws}(a). A close inspection reveals that the CB is used more during the lower price periods; whereas, its usage reduces during high price periods, as depicted in Fig.~\ref{fig:cbwos} and Fig.~\ref{fig:cbws}. Similarly, CGU is used more during high-price periods compared to other local generation units due to its ability to supply electrical and thermal energy simultaneously. As a result, the SB3-MPC provides 17.26\% and 22.46\% cost-efficient scheduling as against heuristic optimization in case study 1 and 2, respectively. BB-MINLP also utilizes local energy sources more during peak-price hours as shown in Fig.~\ref{fig:ESwos}(b) and Fig.~\ref{fig:ESws}(b). Due to under-utilization of the storage units by BB-MINLP in presence of degradation costs reduces the overall cost. The cost-efficiency of BB-MINLP is found to be 2\% more than the proposed SB3-MPC approach with storage unit. However, it can be verified from Fig~!\ref{fig:ESwos} that during low price hours BB-MINLP is unable to utilize the utility grid at its maximum potential. As a consequence, SB3 achieves 3\% more cost-effectiveness over BB-MINLP in absence of storage units.

\begin{table}
\centering
\caption{Change in demand response w.r.t scenarios (in \%)}
{\renewcommand{\arraystretch}{1.0}
\begin{tabular}{|c|c|c|c|c|}
\hline
Scenario & 30 & 40 & 50 & 60 \\ \hline
Case study 1 & $\pm$0.6 & $\pm$0.1 & $\pm$0.4 & $\pm$1.00 \\
Case study 2 & $\pm$1.0 & $\pm$2.0 & $\pm$1.7 & $\pm$0.7 \\ 
Runtime ({\si{\second}}) & 12.48 & 15.97 & 22.04 & 25.48 \\ \hline 
\end{tabular}}
\label{tab:sens}
\end{table}

\subsection{Robustness, Sensitivity Analysis}

To study the MPC's robustness, forecast errors of 2\% and 10\% was introduced. Our study, as recorded in Tab.~\ref{tab:perf}, revealed that with a forecast error of 10\%, the change in savings and PBDR were around 2.5\% and 1.77\% respectively against SB3-based MPC's performance on actual demand. This showed the proposed method's robustness to forecast uncertainties. In addition, PBDR performance of the SB3 is analyzed for both case study 1 and 2, by varying the maximum number of \textit{scenarios}. Tab.~\ref{tab:sens} records the change in PBDR for different number of \textit{scenarios} considering the outcome with 20 \textit{scenarios} as the base performance. Analysis suggests that variation in number of scenarios has an negligible effect on the SB3's performance, whereas the runtime increases significantly with increasing number of scenarios. The iRCGA solver has the capability of population based direct search method, which reduces the dependency of the initial number of \textit{scenarios} in the end result.

\subsection{Computation Performance}
The SB3's computation performance was studied on 10 trials using simulations for 24 {\si{\hour}} duration. The average execution time for different prediction horizons are shown in Tab.~\ref{tab:runtime}. The average execution time of heuristic is found to be 5437 {\si{\second}} when the maximum number of iteration considered is 5000. Therefore, the computational time is higher while solving the problem with receding horizon approach. The approximate time for BB-MINLP with MPC approach to solve the problem is around 720 {\si{\second}} whereas, the computation time of the SB3 is around 10 {\si{\second}} even with a 30 {\si{\hour}} prediction horizon. This shows the SB3-based MPC's capability to be used as a real-time scheduler in place of multi time-step heuristic methods and BB-MINLP solver.
\begin{table}
\centering
\caption{Runtime of the SB3 with various prediction horizon}
{\renewcommand{\arraystretch}{1.0}	
\begin{tabular}{|c|c|c|c|c|c|}
\hline
$N_p$ & 20 & 30 & 40 & 50 & 60 \\ \hline
Runtime ({\si{\second}}) & 7.01 & 7.8 & 8.7 & 9.8 & 10.26 \\ \hline
\end{tabular}}
\label{tab:runtime}
\end{table}

\section{Conclusions}
\label{Sec:Concl}
This investigation presented a novel scheduling algorithm for Multi-Energy Systems (MES) that performs PBDR. The formulation led to a MES scheduling problem with mixed integer nonlinear program (MINLP), a NP-hard problem. To reduce computation complexity, the proposed SB3 solver utilizes the McCormick's bi-linear relaxation and epigraph technique, followed by integrating the simplicity of convex programs and the ability of meta-heuristic optimization to solve complex nonlinear problems. This study also investigated the existence conditions of the lower bound solution of the problem while solved with the SB3 method. The approach was demonstrated on a pilot building in Singapore, a test-bed for MES. Our results showed that the SB3-MPC reduced cost by approximately 17.26\% without storage and 22.46\% with storage when compared with multi-time step dynamic MINLP solved using heuristic optimization. Moreover, SB3-MPC also achieved promising demand response over BB-MINLP solver. Although the proposed method has produced sub-optimal solution compared to BB-MINLP, it showed robustness to forecast errors and has lower computation time significant. Studying the role of market strategies in MES scheduling and extension of this work in multi-objective optimization framework for planning and scheduling are the future course of this investigation. 

\section*{Acknowledgement}
This research is supported by the National Research Foundation, Prime Ministers Office, Singapore under its Energy NIC grant (NRF Award No.: NRF-ENIC-SERTD-SMESNTUJTCI3C-2016).

\bibliographystyle{IEEEtran}
\bibliography{BIB_TII-19-4619}

\begin{IEEEbiography}[{\includegraphics[width=1in,height=1.25in,clip,keepaspectratio]{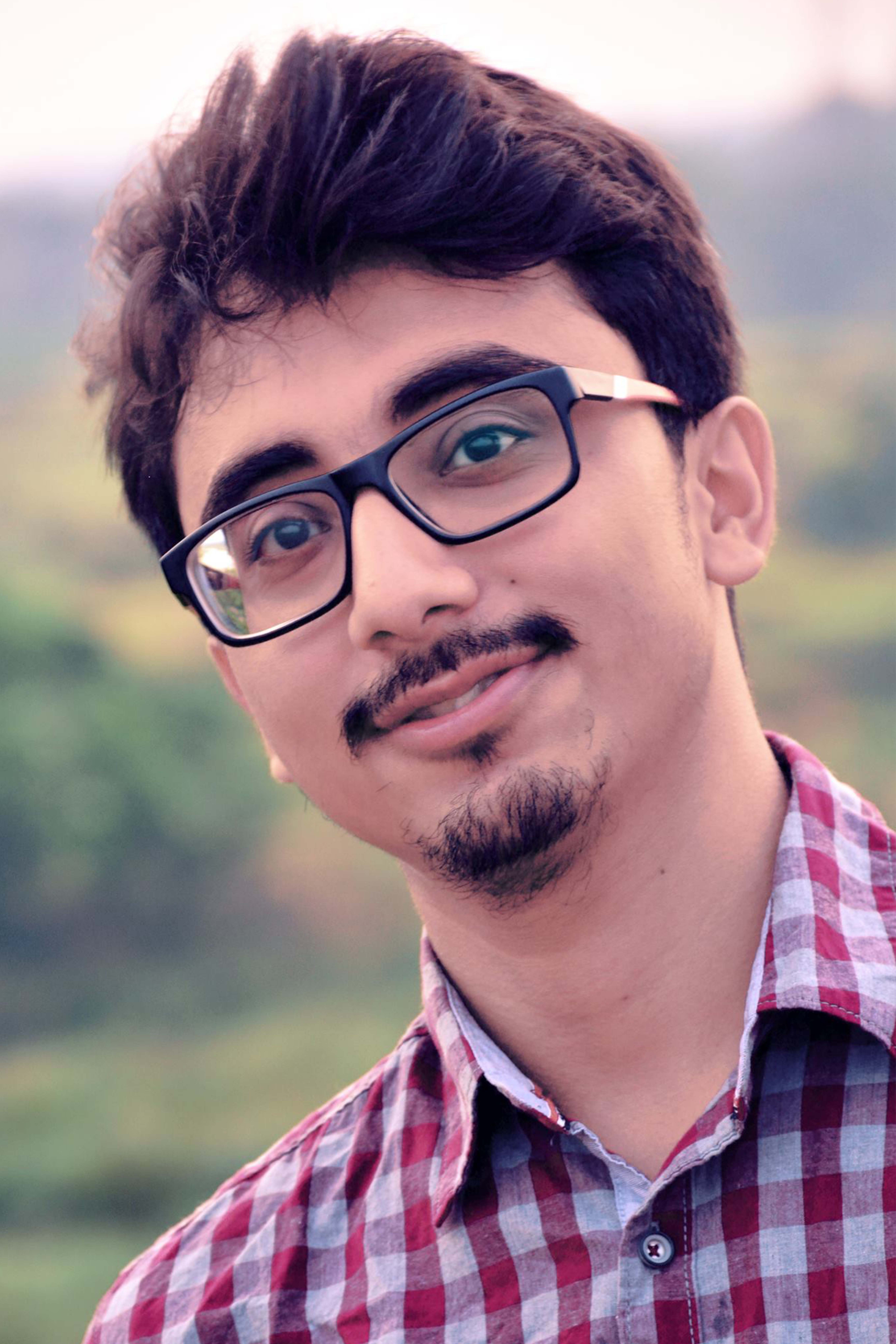}}]{Mainak Dan}
received his bachelors and masters degree in electronics and tele-communication engineering with specialization in control engineering from Jadavpur University in 2014 and 2016, respectively. Currently he is working towards his doctorate in philosophy with Interdisciplinary Graduate Program, Nanyang Technological University, Singapore from 2016. His research interest lies in the problem area of model predictive control, optimal control, mixed integer optimization techniques, energy management systems. 
\end{IEEEbiography}

\begin{IEEEbiography}[{\includegraphics[width=1in,height=1.25in,clip,keepaspectratio]{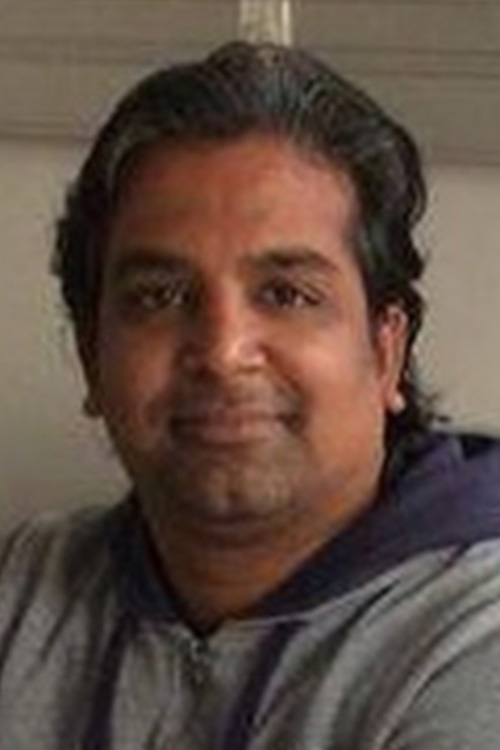}}]{Dr. Seshadhri Srinivasan}
is currently with Berkeley Education Alliance for Research in Singapore, Singapore working on building automation, optimal control, fault-diagnosis, and smart grids. Prior to joining BEARS, he was working with the University of Sannio, Italy, Technical University of Munich, Germany, Tallinn University of Technology, Estonia, and ABB Corporate Research Center at Bangalore. He is a member of the IEEE CSS standing committee on standards.  
\end{IEEEbiography}

\begin{IEEEbiography}[{\includegraphics[width=1in,height=1.25in,clip,keepaspectratio]{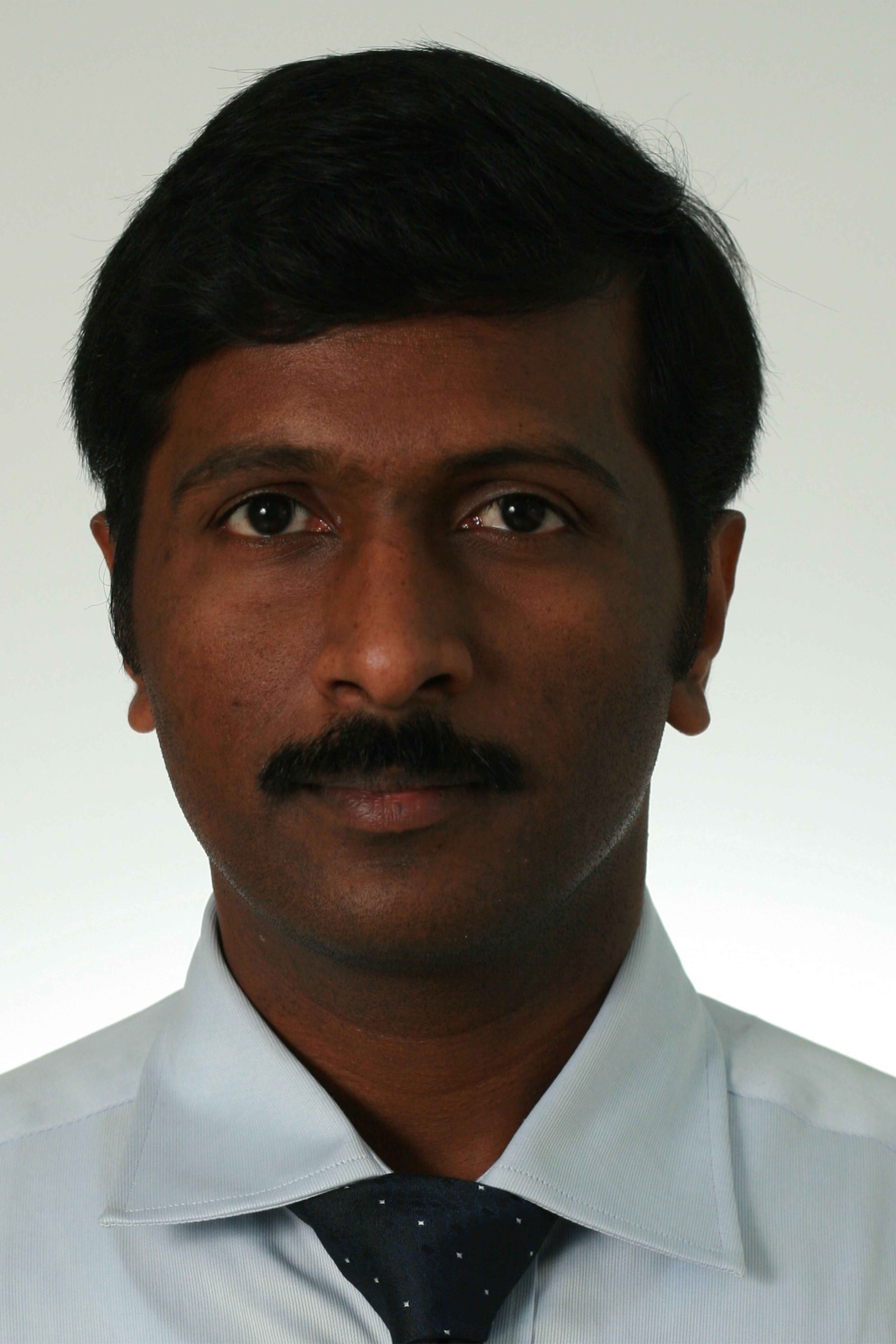}}]{Dr. Suresh Sundaram}
received the B.E degree in electrical and electronics engineering from Bharathiyar University in 1999, and M.E (2001) and Ph.D. (2005) degrees in aerospace engineering from Indian Institute of Science, India. He worked as an Associate Professor with School of Computer Science and Engineering, Nanyang Technological University, Singapore till Oct 2018. Currently he is working as Associate Professor in Department of Aerospace, Indian Institute of Science, Bangalore, India. His research interest includes flight control, unmanned aerial vehicle design, machine learning, applied game theory, optimization, and computer vision.
\end{IEEEbiography}

\begin{IEEEbiography}[{\includegraphics[width=1in,height=1.25in,clip,keepaspectratio]{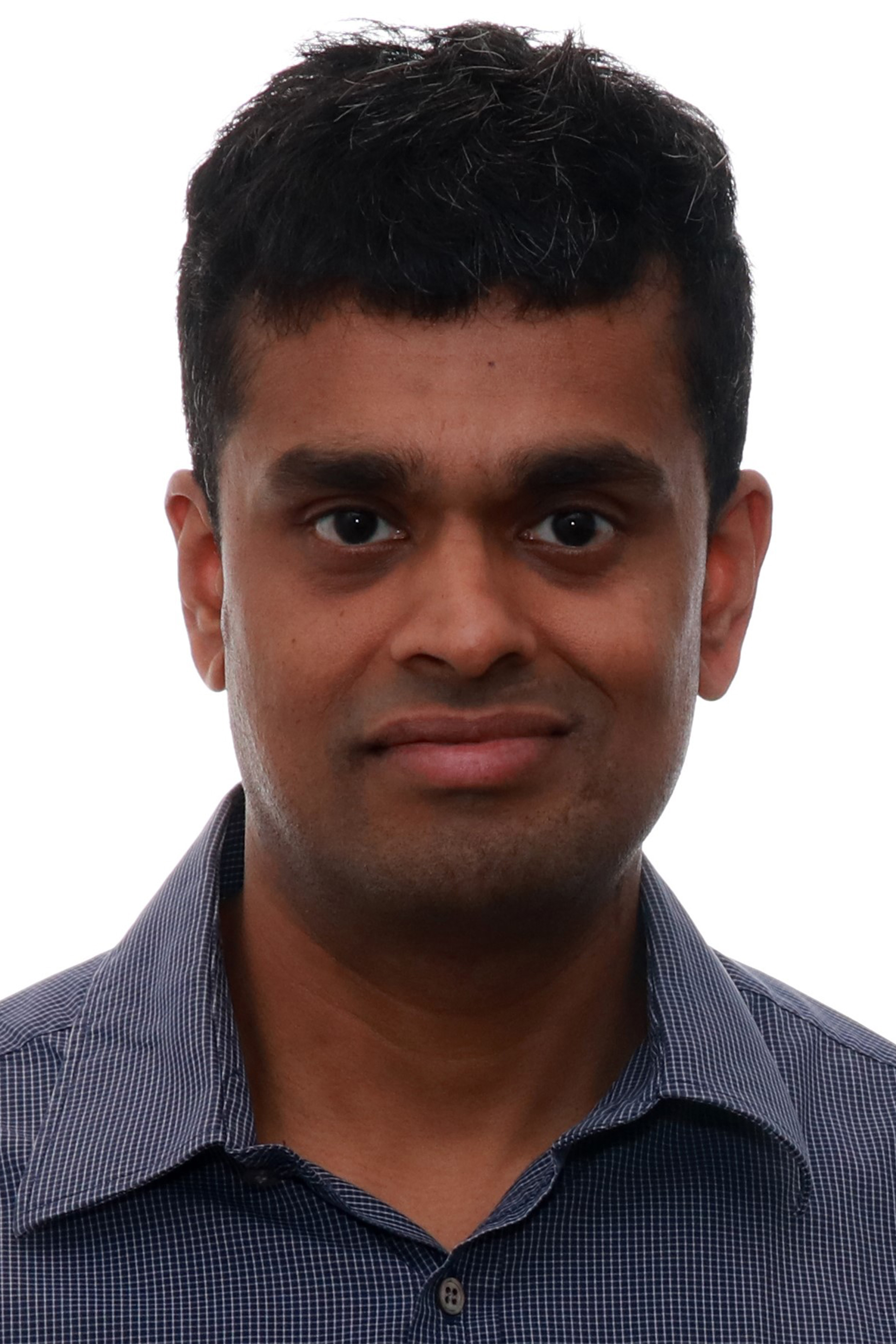}}]{Dr. Arvind Easwaran}
is an Associate Professor in the School of Computer Science and Engineering at Nanyang Technological University (NTU), Singapore. He is also a Cluster Director for the Future Mobility Solutions research programme at the Energy Research Institute in NTU. He received a PhD degree in Computer and Information Science from the University of Pennsylvania, USA, in 2008. In NTU, his research focuses on the design and analysis of real-time and cyber-physical computing systems, including their application in domains such as automotive, manufacturing and urban energy systems.  
\end{IEEEbiography}

\begin{IEEEbiography}[{\includegraphics[width=1in,height=1.25in,clip,keepaspectratio]{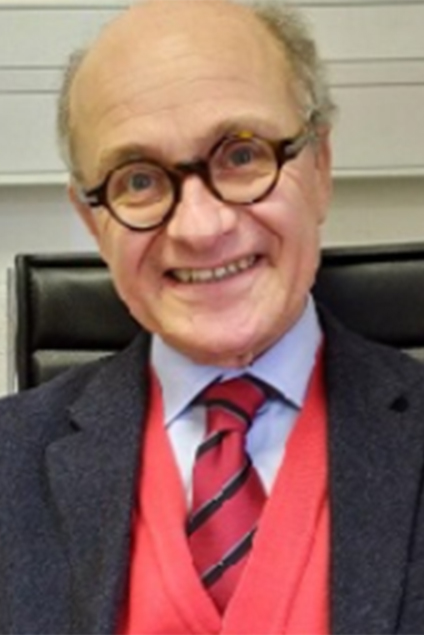}}]{Dr. Luigi Glielmo}
holds a “laurea” degree in Electronic Engineering and a research doctorate in Automatic Control.  He taught at University of Palermo, at University of Naples Federico II and then at University of Sannio at Benevento where he is now a professor of Automatic Control.  His current research interests include singular perturbation methods, model predictive control methods, automotive controls, deep brain stimulation modeling and control, smart-grid control. He seated in the editorial boards of Dynamics and Control and IEEE Transaction on Automatic Control, and has been chair of the IEEE Control Systems Society Technical Committee on Automotive Controls.  He has been head of the Department of Engineering of University of Sannio from 2001 to 2007.
\end{IEEEbiography}

\end{document}